\documentclass[11pt]{article}
\usepackage{a4wide}
\usepackage{cite}
\usepackage{amsmath}
\usepackage{amstext,amsfonts,amsbsy,,amssymb}
\usepackage{amsmath,pifont}
\usepackage{color}
\usepackage{epsfig,psfrag}

\usepackage{,times}
\newfont{\cyr}{wncyr10}
\usepackage{mathrsfs}
\numberwithin{equation}{section}
\renewcommand{\thefootnote}{\fnsymbol{footnote}}

\def\openone{\leavevmode\hbox{\small1\kern-3.8pt\normalsize1}}%
\DeclareMathOperator{\sh}{sh}

\DeclareMathOperator{\sgn}{sgn}

\DeclareMathOperator{\Tr}{Tr}
\DeclareMathOperator{\diag}{diag}

\newcommand{\dphi}{\Delta \phi}

\begin{document}

\baselineskip 17pt
\parskip 7pt

\noindent

\hfill  February 14, 2001

\vspace{24pt}

\begin{center}

  {\Large\textbf{
      The Baxter Equation for Quantum Discrete Boussinesq Equation
      }
    }

  \vspace{24pt}

  {\large Kazuhiro \textsc{Hikami}}
  \footnote[2]{E-mail:
    \texttt{hikami@phys.s.u-tokyo.ac.jp}
    }

  \textsl{Department of Physics, Graduate School of Science,\\
    University of Tokyo,\\
    Hongo 7--3--1, Bunkyo, Tokyo 113--0033, Japan.
    }

(Received: \hspace{40mm})

\end{center}


\begin{center}
  \underline{ABSTRACT}

  \begin{minipage}{14cm}


Studied is the Baxter equation for the quantum discrete Boussinesq
equation.
We explicitly construct the Baxter $\mathcal{Q}$ operator from a
generating function of the local integrals of motion of the affine
Toda lattice field theory, and show that it solves the third order
operator-valued difference equation.

\hfill \texttt{nlin/0102021}

\end{minipage}
\end{center}




\bigskip

\renewcommand{\thefootnote}{\arabic{footnote}}

\section{Introduction}

In analysis of the integrable models,
realized is the importance of the Baxter equation since it was applied
to the eight-vertex
model (XYZ model) to compute the spectrum~\cite{Bax82}.
The Baxter equation is the functional difference
relation among the transfer
matrix $t(\lambda)$
and the auxiliary function $\mathcal{Q}(\lambda)$ which is called the
Baxter $\mathcal{Q}$ operator,
and is powerful enough to determine the
spectrum of the transfer matrix
only from certain  analytic conditions of these functions.

Later Gaudin and Pasquier explicitly constructed
the Baxter $\mathcal{Q}$ operator
in an
integral form for the quantum periodic Toda lattice~\cite{PasqGaud92}.
It was also shown~\cite{PasqGaud92,KuzneSklya98a} that
in the classical limit the
logarithm of the $\mathcal{Q}$ operator is the generating function of
the 
B{\"a}cklund transformation,
\emph{i.e.}, the B{\"a}cklund transformation
is given by the classical limit of the
evolution equation,
$\mathcal{O} \to \mathcal{Q} \, \mathcal{O} \, \mathcal{Q}^{-1}$
(see, for review, Ref.~\citen{Sklyan00b}).
This correspondence suggests that
the  integrable dynamical system on space-time lattice
can be constructed by use of
the $\mathcal{Q}$ operator~\cite{FaddVolk97a,RInouKHikam97c}.
Recent studies indicate that
the $\mathcal{Q}$  operator is useful enough to construct
the
eigenfunction of the Toda lattice~\cite{KharLebe00a}.

Since then the explicit form of the Baxter $\mathcal{Q}$
operator has been derived
for other integrable models such as
the XXX model~\cite{Derka99a,GPronk99a}, the dimer self-trapping
model~\cite{KuznSaleSkly00a}, the quantum KdV
model~\cite{BazhLukyZamo96b}, and the Volterra
model (discrete KdV equation)~\cite{AYVolk96a,HikamRInou98a}.
All these models are governed by the quantum algebra associated to
$s\ell_2$.
In these cases
as was pointed out in  studies of the integrability of the conformal
field theory~\cite{BazhLukyZamo96b}, the Baxter
equation may be related with the universal $\mathcal{R}$ matrix,
$\mathcal{R} \in U_q(\widehat{s\ell}_2) \otimes U_q(\widehat{s\ell}_2)$,
with some quantum  finite-dimensional  representation for the first
$U_q(\widehat{s\ell}_2)$ algebra
and an auxiliary \emph{infinite}-dimensional
$q$-oscillator  representation for the second part.
In fact
proposed was the
universal procedure~\cite{AntonFeigi96a}
to derive  the Baxter equation from the universal $\mathcal{R}$ matrix
by use of the $q$-oscillator representation.
Following this strategy it was shown that   the
$\mathcal{Q}$ operator~\cite{AYVolk96a}
for the Volterra model can be
constructed from the universal $\mathcal{R}$ matrix
for the quantum algebra  $U_q(\widehat{s\ell}_2)$
with the $q$-oscillator representation~\cite{Anton96a}.

In this paper we construct  the
Baxter $\mathcal{Q}$ operator
for the quantum discrete Boussinesq equation which has
$U_q(\widehat{s\ell}_3)$ symmetry.
In the classical limit
the discrete Boussinesq equation is given as
\begin{equation}
  \label{lattice_Boussinesq}
  \begin{split}
    \frac{\mathrm{d} L_n}{\mathrm{d} t}
    & =
    - W_n + W_{n-1}
    + L_n \, \bigl( L_{n+1} - L_{n-1} \bigr) ,
    \\[2mm]
    \frac{\mathrm{d} W_n}{\mathrm{d} t}
    & =
    W_n \,
    \bigl( L_{n+2} - L_{n-1} \bigr) .
  \end{split}
\end{equation}
We stress that $L_n$ and $W_n$ do not mean the Fourier
modes of the generators
$L(x)$ and $W(x)$ but the discretization thereof.
We note
that the higher flow of
eq.~\eqref{lattice_Boussinesq} reduces  in a continuum
limit~\cite{HikamRInou97a} to
\begin{equation}
  \label{usual_Boussinesq}
  \begin{split}
    \frac{\mathrm{d} u}{\mathrm{d} t}
    & =
    u_{xx} - 2 \, w_x ,
    \\[2mm]
    \frac{\mathrm{d} w}{\mathrm{d} t}
    & =
    - w_{xx} 
    + \frac{2}{3} \, u_{xxx}
    + \frac{2}{3} \, u \, u_{x}    .
  \end{split}
\end{equation}
This set of equations gives  the Boussinesq equation
which is generated from the pseudo-differential operator,
$L = \partial^3 + u \, \partial + w$.
In this sense we call
eq.~\eqref{lattice_Boussinesq} as  the  discrete
Boussinesq equation.
The discrete Boussinesq equation~\eqref{lattice_Boussinesq}
was first introduced 
in Ref.~\citen{BeloChal93c} from studies of a discretization of the
$W_3$ algebra,
motivated from the close connection between the soliton equations and
the conformal field theory~\cite{EgYa89,KupeMath89a,RSasaIYama87a}.
The integrability of the discrete Boussinesq equation follows from the
fundamental commutation relation,
\begin{equation}
  \{ \mathbf{T}(x)
  \stackrel{\otimes}{,}
  \mathbf{T}(y) \}
  =
  [ \boldsymbol{r}(x,y) ~,~
  \mathbf{T}(x)
  \otimes  \mathbf{T}(y) ] ,
\end{equation}
where we omit the explicit form of the $\mathbb{Z}_3$ symmetric
classical $\boldsymbol{r}$ matrix.
The monodromy matrix $\mathbf{T}(x)$ is  given from the Lax matrix
as~\cite{HikamRInou97a}
\begin{align}
  \label{Lax_L_W}
  \mathbf{T}(x)
  & =
  \prod_n^\curvearrowleft
  \boldsymbol{L}_n^{\mathrm{W}}(x),
  &
  \boldsymbol{L}_n^{\mathrm{W}}(x)
  &=
  \frac{1}{
    \sqrt[3]{W_n}
    } \,
  \begin{pmatrix}
    x^2 & -x \, L_{n+1} & W_n \\[2mm]
    1 & 0 & 0 \\[2mm]
    0 & 1 & 0
  \end{pmatrix} .
\end{align}
As a result we can construct the integrals of motion from the transfer
matrix
$\Tr \boldsymbol{T}(x)$; some of them are
\begin{align*}
  \mathcal{H}_0
  & = \sum_n \log W_n ,
  \\[2mm]
  \mathcal{H}_1
  & = \sum_n L_n ,
  \\[2mm]
  \mathcal{H}_2
  & =
  \sum_n
  \Bigl(
  \frac{1}{2} \bigl( L_n \bigr)^2
  + L_n \, L_{n+1} - W_n
  \Bigr) .
\end{align*}

As is well known,
the discrete Boussinesq equation has also an intimate relationship with
a lattice analogue of the affine Toda field theory.
We have constructed a generating function $\mathcal{Q}$ of the local
integrals of motion of the  affine
Toda lattice field theory~\cite{Hikam98b}, and have also shown that
this also commute with the transfer matrix of the discrete Boussinesq
equation.
The  function  $\mathcal{Q}$  is constructed
in terms  of the quantum dilogarithm function as the intertwining
operator of the vertex operators.
As was conjectured there,
we shall show  here that this generating function is indeed the Baxter
$\mathcal{Q}$ operator for the discrete Boussinesq equation.

This paper is organized as follows.
In Section~\ref{sec:Boussinesq} we briefly
review the integrable structure of
the quantum discrete Boussinesq equation.
Recalling the relationship with the affine Toda field theory,
we construct the transfer matrix and the generating function
$\mathcal{Q}(\lambda)$ of the local
integrals of motion for the affine Toda lattice field theory.
In Section~\ref{sec:Baxter}
we  explicitly show that the generating function $\mathcal{Q}$ plays a
role of the Baxter $\mathcal{Q}$ operator for the transfer matrix of
the discrete Boussinesq equation.
Based on the duality of the affine Toda lattice field theory,
we show
in Section~\ref{sec:duality}
that we have
the dual Baxter equation for the discrete Boussinesq equation.
In Section~\ref{sec:continue} we briefly comment on the continuum
limit of the discrete Boussinesq equation, and see a relationship
between the previously known results.
In Section~\ref{sec:kp} we consider the generalization for the
$N$-reduced
discrete KP equation which has the quantum group
$U_q(\widehat{s\ell}_N)$ structure.
Last section is devoted for the concluding remarks and discussions.


\section{Quantum Boussinesq Equation}
\label{sec:Boussinesq}

\subsection{Lattice Vertex Operator and Transfer Matrix}

To define the quantization of the discrete Boussinesq
hierarchy~\eqref{Lax_L_W}, we use two sets of the discretized
free fields $\phi_n^{(1)}$ and
$\phi_n^{(2)}$, which  respectively correspond to the Toda fields
associated to the simple roots
$\boldsymbol{\alpha}_1$ and $\boldsymbol{\alpha}_2$.
We use  the free field associated to the maximal root as
\begin{equation}
  \phi_n^{(0)} = - \phi_n^{(1)} - \phi_n^{(2)} .
\end{equation}
The commutation relations are defined as a discrete analogue of the
usual relations of the free chiral  fields by\footnote{
  We note that the dynamical variables $\alpha_n$ and $\beta_n$
  in Ref.~\citen{Hikam97c}
  are defined by
  \begin{align*}
    \alpha_n & = \mathrm{e}^{\dphi_n^{(1)} - \mathrm{i} \gamma} ,
    &
    \beta_n & = \mathrm{e}^{\dphi_n^{(1)} + \dphi_n^{(2)} - \mathrm{i}
      \gamma} ,
  \end{align*}
  with
  $q  = \mathrm{e}^{\mathrm{i} \gamma}$.
  }
\begin{equation}
  \label{commutation_phi}
  \begin{split}
    [ \phi_m^{(a)} ~,~ \phi_n^{(a)} ]
    & = 2 \, \mathrm{i} \, \gamma ,
    \quad
    \text{for $m>n$, $a=1,2$} ,
    \\[2mm]
    [ \phi_n^{(1)} ~,~ \phi_m^{(2)} ]
    & =
    \begin{cases}
      \mathrm{i} \, \gamma , & \text{for $n \leq m$},
      \\
      - \mathrm{i} \, \gamma , & \text{for $n >m$} .
    \end{cases}
  \end{split}
\end{equation}
Throughout this paper we set the deformation parameter $q$ as
\begin{equation}
  q=\mathrm{e}^{\mathrm{i} \gamma} .
\end{equation}
Hereafter for our convention
we often use a difference of the lattice free field
\begin{equation}
  \label{define_dphi}
  \dphi_n^{(a)} = \phi_n^{(a)} - \phi_{n+1}^{(a)},
\end{equation}
whose commutation relations become local.

With these settings, we  define the discrete screening charges as in
the case of the conformal field theory (see
e.g. Ref.~\citen{DiFrMathSene97});
\begin{equation}
  Q^{(a)} = \sum_n \mathrm{e}^{\phi_n^{(a)}} ,
\end{equation}
for $a=0,1,2$.
We can see  that  the Serre relation of the algebra
$U_q(\widehat{s\ell}_3)$
is fulfilled;
\begin{equation}
  \label{Serre}
  \bigl( Q^{(a)} \bigr)^2 \, Q^{(b)}
  -(q+q^{-1}) \cdot
  Q^{(a)} \, Q^{(b)} \, Q^{(a)}
  +
  Q^{(b)} \,
  \bigl( Q^{(a)} \bigr)^2
  = 0 ,
\end{equation}
for $a \neq b$.
The generators of the quantum $W_3$ algebra are then 
defined so as to
commute with these  screening charges.
Note that the dual screening charges as a summation of the dual vertex
operators,
\begin{equation}
  \Tilde{Q}^{(a)}
  =
  \sum_n
  \mathrm{e}^{\frac{\pi}{\gamma} \phi_n^{(a)}} ,
\end{equation}
commute with $Q^{(a)}$, and that they
satisfy  the same  Serre relation~\eqref{Serre} replacing $q$ with
\begin{equation}
  \Tilde{q}= \mathrm{e}^{\mathrm{i} \frac{\pi^2}{\gamma}} .
\end{equation}

As was   shown in Ref.~\citen{HikamRInou97a}, the Lax
matrix~\eqref{Lax_L_W} for the discrete Boussinesq equation
is gauge-equivalent with
\begin{align}
  \label{Lax_1}
  \mathbf{L}_n(\lambda)
  & =
  g(\lambda) \,
  \begin{pmatrix}
    \mathrm{e}^{\frac{1}{3}\dphi_n^{(1)} - \frac{1}{3} \dphi_n^{(2)}}
    &&\\
    & \mathrm{e}^{\frac{1}{3}\dphi_n^{(1)} + \frac{2}{3}
      \dphi_n^{(2)}}
    &\\
    && \mathrm{e}^{-\frac{2}{3}\dphi_n^{(1)} - \frac{1}{3} \dphi_n^{(2)}}
  \end{pmatrix}
  \cdot
  \begin{pmatrix}
    \mathrm{e}^\lambda & \mathrm{e}^{-\lambda} & 1 \\[2mm]
    1 & \mathrm{e}^\lambda & \mathrm{e}^{-\lambda} \\[2mm]
    \mathrm{e}^{-\lambda} & 1  &\mathrm{e}^\lambda 
  \end{pmatrix}
  \\
  & \equiv
  g(\lambda) \cdot
  \mathbf{C}_n \cdot \mathbf{X}(\lambda) ,
  \nonumber
\end{align}
where we have modified  the spectral parameter  by
$x^{-3} =  1 - \mathrm{e}^{- 3 \lambda}$ from eq.~\eqref{Lax_L_W}.
The  function $g(\lambda)$ is a normalization function which will be
defined
below.
We   introduce  diagonal elements of the matrix $\mathbf{C}_n$ as
\begin{equation}
  \mathbf{C}_n = \diag
  \Bigl(
  c_n^{(1)} , c_n^{(2)} , c_n^{(3)}
  \Bigr),
\end{equation}
which corresponds to the weights  in the vector representation
$\boldsymbol{3}$ of $s\ell_3$.
Following
the standard way of studies in the integrable systems, we define the
monodromy matrix and the transfer matrix as
\begin{align}
  \label{monodromy_3}
  \mathbf{T}(\lambda)
  &=
   \prod_n^\curvearrowleft \mathbf{L}_n(\lambda) ,
  &
  t_1(\lambda)
  &=
  \Tr \mathbf{T}(\lambda) .
\end{align}
The monodromy matrix satisfies
the Yang--Baxter equation~\cite{Hikam97c},
\begin{gather}
  \label{RTT_1}
  {\mathbf{R}}(\lambda - \mu) \cdot
  \overset{1}{\mathbf{T}}(\lambda) 
  \,
  \overset{2}{\mathbf{T}}(\mu )
  =
  \overset{2}{\mathbf{T}}(\mu) 
  \,
  \overset{1}{\mathbf{T}}(\lambda )
  \cdot
  {\mathbf{R}}(\lambda - \mu) .
\end{gather}
Here the $\mathbf{R}$-matrix is $\mathbb{Z}_3 \otimes \mathbb{Z}_3$
symmetric, and reads as
\begin{equation}
  {\mathbf{R}}(\lambda)
  =
  \left(
    \begin{array}{ccc|ccc|ccc}
      a &&&&&&&&\\
      & b && c &&&&&\\
      && \Bar{b} &&&& \Bar{c} &&\\
      \hline
      & \Bar{c} && \Bar{b} &&&&&\\
      &&&& a &&&&\\
      &&&&& b && c &\\
      \hline
      && c &&&& b &&\\
      &&&&& \Bar{c} && \Bar{b} &\\
      &&&&&&&& a
    \end{array}
  \right)  ,
\end{equation}
with
\begin{align*}
  a & = \sh(\frac{3}{2} \, \lambda - \frac{1}{2}\, \mathrm{i}\, \gamma) ,
  \\
  b
  & =
  \mathrm{e}^{  \frac{1}{6} \mathrm{i} \gamma}
  \sh ( \frac{3}{2} \, \lambda ) ,
  &
  \Bar{b}
  & =
  \mathrm{e}^{ - \frac{1}{6} \mathrm{i} \gamma}
  \sh ( \frac{3}{2} \, \lambda ) ,
  \\[2mm]
  c
  & =
  \mathrm{e}^{ - \frac{1}{2}\lambda}
  \sh ( - \frac{1}{2} \, \mathrm{i} \, \gamma) ,
  &
  \Bar{c}
  & =
  \mathrm{e}^{\frac{1}{2}\lambda}
  \sh ( - \frac{1}{2} \, \mathrm{i} \, \gamma) .
\end{align*}
It follows that the transfer matrix commute with each other,
\begin{equation}
  [ t_1(\lambda ) ~,~ t_1(\mu) ] = 0 ,
\end{equation}
which supports  the quantum integrability of our system.

We note that  under the quasi-periodic boundary condition,
\begin{equation}
  \label{lattice_boundary}
  \phi_{n+L}^{(a)}
  =
  \phi_n^{(a)}
  +
  L \cdot \mathrm{i} \, P^{(a)} ,
  \qquad
  \text{for $a=1,2$},
\end{equation}
the transfer matrix~\eqref{monodromy_3} can be rewritten as
\begin{equation}
  \label{rewrite_t1}
  t_1(\lambda)
  =
  \Tr
  \biggl(
  \mathrm{e}^{
    - L \, \mathrm{i} \,
    \bigl(
    - P^{(1)} \mathbf{h}_3
    +
    P^{(2)} \mathbf{h}_2
    \bigr)
    }
  \cdot
  \overset{\curvearrowleft}{\prod_{n=1}^{{L}}}
  g(\lambda) \, \mathrm{e}^{\frac{1}{3} \mathrm{i} \gamma} \,
  \mathbf{D}_n \,
  \mathbf{X}(\lambda) \,
  \mathbf{D}_n^{~-1} 
  \biggr) ,
\end{equation}
where $3 \times 3$  diagonal matrix $\mathbf{D}_n$ is given by
\begin{gather*}
  \mathbf{D}_n
  =
  \exp
  \Bigl(
  - \phi_n^{(1)} \, \mathbf{h}_3
  +
  \phi_n^{(2)} \, \mathbf{h}_2
  \Bigr),
  \\[2mm]
%
%
  \mathbf{h}_a
  = \mathbf{E}_{a a} - \frac{1}{3} \, \openone .
%
\end{gather*}
A  matrix $\mathbf{E}_{ij}$  is
$\bigl( \mathbf{E}_{ij} \bigr)_{mn} = \delta_{i,m} \, \delta_{j,n}$.

For our purpose to construct the Baxter equation for the discrete Boussinesq
equation,
we use another
transfer matrix which is associated with
a representation
$\boldsymbol{3}^*$.
Generally
the  monodromy matrix for this type of
representation can be constructed from
one for the vector representation
by the
fusion method~\cite{KuliReshSkly81a},
and   in the $s\ell_3$ case the Lax matrix is simply
defined from
$\overline{\mathbf{L}}_n(\lambda)
\propto
\Bigl( \bigl(
\mathbf{L}_n(\lambda)
\bigr)^{-1}\Bigr)^{\mathrm{t}}$,
where $\mathrm{t}$ denotes a transpose~\cite{KuliResh86};
\begin{align}
  \overline{\mathbf{L}}_n(\lambda)
  & =
  \overline{g}(\lambda) \,
  \begin{pmatrix}
    \mathrm{e}^{-\frac{1}{3}\dphi_n^{(1)} + \frac{1}{3} \dphi_n^{(2)}}
    &&\\
    & \mathrm{e}^{-\frac{1}{3}\dphi_n^{(1)} - \frac{2}{3}
      \dphi_n^{(2)}}
    &\\
    && \mathrm{e}^{\frac{2}{3}\dphi_n^{(1)} + \frac{1}{3} \dphi_n^{(2)}}
  \end{pmatrix}
  \cdot
  \begin{pmatrix}
    \mathrm{e}^\lambda & -1 & 0 \\[2mm]
    0 &\mathrm{e}^\lambda & -1  \\[2mm]
    -1 & 0 & \mathrm{e}^\lambda 
  \end{pmatrix}
  \\
  & \equiv
  \overline{g}(\lambda) \cdot
  \mathbf{C}_n^{~-1} \cdot
  \overline{\mathbf{X}}(\lambda) ,
  \nonumber 
\end{align}
where as before
$\overline{g}(\lambda)$ is a normalization function.
One  sees
that
each element of $\mathbf{C}_n^{~-1}$ corresponds to weights of
$\boldsymbol{3}^*$.
Correspondingly the monodromy matrix and the transfer matrix for this
representation is respectively defined by
\begin{align}
  \overline{\mathbf{T}}(\lambda)
  &=
  \prod_n^\curvearrowleft \overline{\mathbf{L}}_n(\lambda) ,
  &
  t_2(\lambda)
  & =
  \Tr \overline{\mathbf{T}}(\lambda) .
  \label{transfer_adjoint}
\end{align}
We note that under the boundary condition~\eqref{lattice_boundary}
the transfer matrix  $t_2(\lambda)$  has a form,
\begin{equation}
  \label{rewrite_t2}
  t_2(\lambda)
  =
  \Tr
  \biggl(
  \mathrm{e}^{
    L \, \mathrm{i} \,
    \bigl(
    - P^{(1)} \mathbf{h}_3
    +
    P^{(2)} \mathbf{h}_2
    \bigr)
    }
  \cdot
  \overset{\curvearrowleft}{\prod_{n=1}^{{L}}}
  \overline{g}(\lambda) \,
  \mathrm{e}^{\frac{1}{3} \mathrm{i} \gamma} \,
  \mathbf{D}_n^{~-1} \,
  \overline{\mathbf{X}}(\lambda) \,
  \mathbf{D}_n
  \biggr) ,
\end{equation}

The commutativity of this transfer matrix follows from
eq.~\eqref{RTT_1} as follows.
We have the fundamental commutation relation for
$\overline{\mathbf{T}}(\lambda)$ from eq.~\eqref{RTT_1} as
\begin{gather}
  \Bigl(  \bigl(
  {\mathbf{R}}(\lambda - \mu) \cdot
  \bigr)^{-1} \Bigr)^{\mathrm{t}_1 \mathrm{t}_2}
  \,
  \overset{1}{\overline{\mathbf{T}}}(\lambda) 
  \,
  \overset{2}{\overline{\mathbf{T}}}(\mu )
  =
  \overset{2}{\overline{\mathbf{T}}}(\mu) 
  \,
  \overset{1}{\overline{\mathbf{T}}}(\lambda )
  \,
  \Bigl( \bigl(
  {\mathbf{R}}(\lambda - \mu)
  \bigr)^{-1} \Bigr)^{\mathrm{t}_1 \mathrm{t}_2} ,
\end{gather}
where $\mathrm{t}_a$ denotes a transpose over the $a$-th space.
We see the commutativity of the transfer matrix;
\begin{equation}
  [ t_2(\lambda ) ~,~ t_2(\mu) ] = 0 .
\end{equation}
Furthermore eq.~\eqref{RTT_1} also gives
\begin{equation}
  \Bigl(  \bigl(
  \mathbf{R}(\lambda - \mu)
  \bigr)^{\mathrm{t}_2} \Bigr)^{-1}
  \cdot
  \overset{1}{\mathbf{T}}(\lambda) \
  \overset{2}{\overline{\mathbf{T}}}(\mu)
  =
  \overset{2}{\overline{\mathbf{T}}}(\mu) \
  \overset{1}{\mathbf{T}}(\lambda) \cdot
  \Bigl(  \bigl(
  \mathbf{R}(\lambda - \mu)
  \bigr)^{\mathrm{t}_2} \Bigr)^{-1} ,
\end{equation}
which proves the commutativity among two transfer matrices,
\begin{equation}
  [t_1(\lambda) ~,~ t_2(\mu) ] = 0 .
\end{equation}

\subsection{Fundamental $\mathcal{L}$  Operator}

We shall construct the generating function of the local integrals of
motion for the affine Toda lattice field theory.

For the Heisenberg operators $\Hat{p}$ and $\Hat{q}$ satisfying
\begin{equation*}
  [ \Hat{p} ~,~ \Hat{q} ] = - 2 \, \mathrm{i} \, \gamma ,
\end{equation*}
we define the fundamental
$\Hat{\mathcal{L}}$ operator as~\cite{Hikam97c}\footnote{
  Strictly speaking, the operator~\eqref{introduce_L} is a non compact
  analogue of that was introduced in Ref.~\citen{Hikam97c},
  \emph{i.e.}, we replace the $q$-exponential
  function~\eqref{q_exponential} with the Faddeev integral
  integral~\eqref{define_Phi}~\cite{Hikam00b}.
}
\begin{equation}
  \label{introduce_L}
  \Hat{\mathcal{L}}(\lambda ;  \Hat{q} , \Hat{p})
  =
  \frac{1}{\Phi_\gamma(\lambda + \Hat{p})}
  \cdot
  \frac{1}{\Phi_\gamma(2 \, \lambda + \Hat{q})}
  \cdot
  \frac{\Theta_\gamma( \Hat{q})}
  {\Phi_\gamma(\lambda -  \Hat{q})}
  \cdot
  \frac{\Theta_\gamma( \Hat{p})}
  {\Phi_\gamma(\lambda -  \Hat{p})}  .
\end{equation}
Here $\lambda$ is a spectral parameter, and commute with every quantum
operators, and the functions
$\Phi_\gamma(\varphi)$ and $\Theta_\gamma(\varphi)$ are respectively
defined by
\begin{align}
  \label{define_Phi}
  \Phi_\gamma(\varphi)
  & =
  \exp
  \left(
    \int_{\mathbb{R} + \mathrm{i} \, 0}
    \frac{
      \mathrm{e}^{- \mathrm{i} \varphi x}
      }{
      4 \, \sh(\gamma x) \, \sh(\pi x)
      }
    \frac{\mathrm{d}x}{x}
  \right) ,
  \\[2mm]
  \label{introduce_Theta}
  \Theta_\gamma(\varphi)
  & =
  \Phi_\gamma( \varphi) \cdot \Phi_\gamma( - \varphi) .
\end{align}
%
The integral $\Phi_\gamma(\varphi)$ was first introduced by
Faddeev~\cite{LFadd95a} as a non-compact version of the quantum
dilogarithm function.
See Appendix~\ref{sec:dilog} for properties of these functions,
and we
especially recall that
the function $\Theta_\gamma(\varphi)$ is the
Gaussian~\eqref{define_Theta}.
By use of the difference equation~\eqref{difference_Phi}, we
have~\cite{Hikam97c}
\begin{gather}
  \label{formula_for_F}
  \begin{split}
    \Hat{\mathcal{L}}(\lambda ; \Hat{q} - 2 \, \mathrm{i} \, \gamma,
    \Hat{p})
    & =
    \Hat{\mathcal{L}}(\lambda ; \Hat{q} , \Hat{p})
    \cdot
    \bigl(
    \mathrm{e}^\lambda
    + \mathrm{e}^{-\lambda - \mathrm{i} \gamma + \Hat{q}}
    + \mathrm{e}^{- \mathrm{i} \gamma + \Hat{p}}
    \bigr)
    \cdot
    \frac{1}{
      1
      + \mathrm{e}^{\lambda - \mathrm{i} \gamma + \Hat{q}}
      + \mathrm{e}^{-\lambda - \mathrm{i} \gamma + \Hat{p}}
      }
    \\
    & =
    \frac{1}{
      \mathrm{e}^{-\lambda}
      + \mathrm{e}^{\lambda-\mathrm{i}\gamma+\Hat{q}}
      + \mathrm{e}^{\mathrm{i}\gamma+\Hat{p}}
      }
    \cdot
    \bigl(
    1
    + \mathrm{e}^{-\lambda-\mathrm{i}\gamma+\Hat{q}}
      + \mathrm{e}^{\lambda+\mathrm{i}\gamma+\Hat{p}}
    \bigr) \cdot
    \Hat{\mathcal{L}}(\lambda ; \Hat{q} , \Hat{p}) ,
  \end{split}
  \\[3mm]
  \begin{split}
    \Hat{\mathcal{L}}(\lambda ; \Hat{q}, \Hat{p} - 2 \, \mathrm{i}\, \gamma)
    & =
    \Hat{\mathcal{L}}(\lambda ; \Hat{q} , \Hat{p})
    \cdot
    \bigl(
    1
    + \mathrm{e}^{\lambda + \mathrm{i} \gamma + \Hat{q}}
    + \mathrm{e}^{-\lambda - \mathrm{i} \gamma + \Hat{p}}
    \bigr)
    \cdot
    \frac{1}{
      \mathrm{e}^{-\lambda}
      + \mathrm{e}^{\mathrm{i} \gamma + \Hat{q}}
      + \mathrm{e}^{\lambda - \mathrm{i} \gamma + \Hat{p}}
      }
    \\
    & =
    \frac{1}{
      1
      + \mathrm{e}^{-\lambda-\mathrm{i}\gamma+\Hat{q}}
      + \mathrm{e}^{\lambda - \mathrm{i}\gamma+\Hat{p}}
      }
    \cdot
    \bigl(
    \mathrm{e}^{\lambda}
    + \mathrm{e}^{-\mathrm{i}\gamma+\Hat{q}}
      + \mathrm{e}^{-\lambda-\mathrm{i}\gamma+\Hat{p}}
    \bigr) \cdot
    \Hat{\mathcal{L}}(\lambda ; \Hat{q} , \Hat{p}) .
  \end{split}
\end{gather}

As two operators $\dphi_n^{(1)}$ and $\dphi_n^{(2)}$ are the
Heisenberg operators,
$[ \dphi_n^{(1)} ~,~ \dphi_n^{(2)} ] = 2 \, \mathrm{i} \, \gamma$,
we denote for brevity
\begin{equation}
  \label{fundamental_L}
  \mathcal{L}_n(\lambda ; \phi)
  =
  \Hat{\mathcal{L}}(\lambda ; \dphi_n^{(1)} , \dphi_n^{(1)} + \dphi_n^{(2)})
  .
\end{equation}
We collect useful properties of this operator in
Appendix~\ref{sec:operator}.
This  operator acts as the intertwining operator for the vertex
operators satisfying~\cite{Hikam98b}
\begin{gather}
  \label{intertwine_vertex}
  \Bigl(
  \mathrm{e}^{\phi_n^{(a)}}
  + \mathrm{e}^\lambda \, \mathrm{e}^{\phi_{n+1}^{(a)}}
  \Bigr) \cdot
  \mathcal{L}_n(\lambda ;  \phi)
  =
  \mathcal{L}_n(\lambda ;  \phi)
  \cdot
  \Bigl(
  \mathrm{e}^\lambda \, \mathrm{e}^{\phi_n^{(a)}}
  +  \mathrm{e}^{\phi_{n+1}^{(a)}}
  \Bigr) ,
\end{gather}
for $a=0,1,2$.
The duality~\eqref{duality_Phi}
of the integral~\eqref{define_Phi}  indicates that
the operator
$\mathcal{L}_n(\lambda ; \phi)$ also intertwines the dual vertex operator;
\begin{gather}
  \Bigl(
  \mathrm{e}^{\frac{\pi}{\gamma} \phi_n^{(a)}}
  + \mathrm{e}^{\frac{\pi}{\gamma} \lambda}  \,
  \mathrm{e}^{\frac{\pi}{\gamma} \phi_{n+1}^{(a)}}
  \Bigr) \cdot
  \mathcal{L}_n(\lambda ;  \phi )
  =
  \mathcal{L}_n(\lambda ;  \phi )
  \cdot
  \Bigl(
  \mathrm{e}^{\frac{\pi}{\gamma} \lambda} \,
  \mathrm{e}^{\frac{\pi}{\gamma} \phi_n^{(a)}}
  +  \mathrm{e}^{\frac{\pi}{\gamma} \phi_{n+1}^{(a)}}
  \Bigr) .
  \label{dual_intertwine}
\end{gather}
What is important here is that the fundamental $\mathcal{L}$
operator~\eqref{fundamental_L}
itself satisfies the Yang--Baxter equation~\cite{Hikam97d},
\begin{equation}
  \label{YBE_for_L}
  \mathcal{L}_n(\lambda ; \phi) \,
  \mathcal{L}_{n+1}(\lambda +\mu; \phi) \,
  \mathcal{L}_n(\mu ; \phi)
  =
  \mathcal{L}_{n+1}(\mu ; \phi) \,
  \mathcal{L}_n(\lambda + \mu ; \phi) \,
  \mathcal{L}_{n+1}(\lambda ; \phi)  .
\end{equation}
Thus when we set the $\mathcal{Q}$ operator as
\begin{equation}
  \label{Q_operator}
  \mathcal{Q}(\lambda)
  =
  \prod_n^\curvearrowleft
  f(\lambda) \cdot
  \mathcal{L}_n(\lambda ; \phi) ,
\end{equation}
with a normalization   $f(\lambda)$,
we see that they commute between themselves with arbitrary spectral parameters;
\begin{equation}
  [ \mathcal{Q}(\lambda ) ~,~ \mathcal{Q}(\mu) ]
  = 0 .
\end{equation}
Furthermore
the fundamental $\mathcal{L}$
operator also
satisfies~\cite{Hikam97c}
\begin{equation}
  \mathcal{L}_n( \lambda - \mu ; \phi) \,
  \mathbf{L}_{n+1}(\lambda) \,
  \mathbf{L}_n (\mu)
  =
  \mathbf{L}_{n+1}(\mu) \,
  \mathbf{L}_n(\lambda) \,
  \mathcal{L}_n(\lambda - \mu ; \phi) ,
\end{equation}
which proves
 that the transfer matrices $t_1(\lambda)$ and
$t_2(\lambda)$ commute with the $\mathcal{Q}$
operator~\eqref{Q_operator},
\begin{equation}
  \label{Q_commute_t}
  [\mathcal{Q}(\lambda) ~,~ t_1(\mu) ]
  =
  [\mathcal{Q}(\lambda) ~,~ t_2(\mu) ]
  = 0 .
\end{equation}
In conclusion we have a commutative family;
the transfer matrices $t_1(\lambda)$ and $t_2(\lambda)$, and the
operator $\mathcal{Q}(\lambda)$.
We note that
intertwining relations~\eqref{intertwine_vertex}
and~\eqref{dual_intertwine}
proves the commutativity between the $\mathcal{Q}$ operator and the
screening charges,
\begin{equation}
  [ \mathcal{Q}(\lambda) ~,~ Q^{(a)} ]
  =
  [ \mathcal{Q}(\lambda) ~,~ \Tilde{Q}^{(a)} ]
  = 0  .
\end{equation}
As a consequence,
the $\mathcal{Q}$ operator is a generating function of
the local integrals of motion of  the affine $\widehat{s\ell_3}$ Toda
field theory~\cite{Hikam98b}, whose Hamiltonian and its dual are
\begin{align*}
  \mathcal{H}   
  & =
  \sum_{a=0}^2 Q^{(a)} ,
  &
  \Tilde{\mathcal{H}}   
  & =
  \sum_{a=0}^2 \Tilde{Q}^{(a)}  .
\end{align*}
See Refs.~\citen{EnriqFeigi95,Kryuk95a} for the geometrical
interpretation of the screening charges in the classical limit of the
sine--Gordon system.

\section{Baxter Equation}
\label{sec:Baxter}


We shall  show that the $\mathcal{Q}$ operator defined in
eq.~\eqref{Q_operator}
plays a role of
{the Baxter $\mathcal{Q}$ operator}
for the transfer matrices~\eqref{monodromy_3}
and~\eqref{transfer_adjoint} of the quantum discrete Boussinesq
equation.

\paragraph{\underline{Step 0}}
We prepare  some formulae of
the fundamental $\mathcal{L}$ operator~\eqref{fundamental_L};
using elements of matrix
$\mathbf{C}_n = \diag(c_n^{(1)} , c_n^{(2)} , c_n^{(3)})$ in
eq.~\eqref{Lax_1},
we have
\begin{gather}
  \label{L_formula_1}
  \begin{split}
    \mathrm{e}^{-\lambda - \frac{1}{3}  \mathrm{i}  \gamma}
    \cdot
    \mathcal{L}_n(\lambda+\frac{2}{3} \,  \mathrm{i} \, \gamma ; \phi)
    & =
    \mathcal{L}_n^{(1)}(\lambda ; \phi)
    \cdot
    \Bigl(
    c_n^{(1)} \, \mathrm{e}^\lambda + c_n^{(2)} \, \mathrm{e}^{-\lambda}
    + c_n^{(3)}
    \Bigr)
    \\
    &=
    \mathcal{L}_n^{(2)}(\lambda ; \phi)
    \cdot
    \Bigl(
    c_n^{(1)}  + c_n^{(2)} \, \mathrm{e}^\lambda
    + c_n^{(3)} \, \mathrm{e}^{-\lambda}
    \Bigr)
    \\
    & =
    \mathcal{L}_n^{(3)}(\lambda ; \phi)
    \cdot
    \Bigl(
    c_n^{(1)} \, \mathrm{e}^{-\lambda} + c_n^{(2)} 
    + c_n^{(3)} \, \mathrm{e}^\lambda
    \Bigr) ,
  \end{split}
  \displaybreak[0]
  \\[3mm]
  \label{L_formula_2}
  \begin{split}
    \mathrm{e}^{-\frac{1}{3}\mathrm{i} \gamma}
    (\mathrm{e}^{3 \lambda} - 1)  \cdot
    \mathcal{L}_n(\lambda - \frac{2}{3} \,  \mathrm{i} \, \gamma ;\phi)
    & =
    \Bigl( c_n^{(2)} \Bigr)^{-1} \cdot
    \left(
      \mathcal{L}_n^{(1)}(\lambda ; \phi) \cdot \mathrm{e}^\lambda
      - \mathcal{L}_n^{(2)} (\lambda ; \phi)
    \right)
    \\
    & =
    \Bigl( c_n^{(3)} \Bigr)^{-1} \cdot
    \left(
      \mathcal{L}_n^{(2)}(\lambda ; \phi) \cdot \mathrm{e}^\lambda
      - \mathcal{L}_n^{(3)} (\lambda ; \phi)
    \right) 
    \\
    & =
    \Bigl( c_n^{(1)} \Bigr)^{-1} \cdot
    \left(
      \mathcal{L}_n^{(3)}(\lambda ; \phi) \cdot \mathrm{e}^\lambda
      - \mathcal{L}_n^{(1)} (\lambda ; \phi)
    \right) .
  \end{split}
\end{gather}
Here we  have defined $\mathcal{L}_n^{(a)}(\lambda; \phi)$ by
\begin{equation*}
  \mathcal{L}_n^{(a)}(\lambda ; \phi)
  =
  \Bigl( c_{n+1}^{(a)} \Bigr)^{-1} \cdot
  \mathcal{L}_n(\lambda ; \phi) \cdot
  c_{n+1}^{(a)} ,
\end{equation*}
which, using the commutation relations~\eqref{commutation_phi},
are  explicitly written as
\begin{align*}
  \mathcal{L}_n^{(1)}(\lambda ; \phi)
  & =
  \Hat{\mathcal{L}}(\lambda ; \dphi_n^{(1)} - \frac{2}{3} \, \mathrm{i} \, \gamma ,
  \dphi_n^{(1)} + \dphi_n^{(2)} + \frac{2}{3} \, \mathrm{i} \, \gamma)
  ,
  \\[2mm]
  \mathcal{L}_n^{(2)}(\lambda ; \phi)
  & =
  \Hat{\mathcal{L}}(\lambda ;
  \dphi_n^{(1)} - \frac{2}{3} \, \mathrm{i} \, \gamma ,
  \dphi_n^{(1)} + \dphi_n^{(2)} - \frac{4}{3} \,  \mathrm{i} \,
  \gamma) ,
  \\[2mm]
  \mathcal{L}_n^{(3)}(\lambda ; \phi)
  & =
  \Hat{\mathcal{L}}(\lambda ; \dphi_n^{(1)} + \frac{4}{3} \,
  \mathrm{i} \, \gamma ,
  \dphi_n^{(1)} + \dphi_n^{(2)} + \frac{2}{3} \, \mathrm{i} \, \gamma)
  .
\end{align*}
See Appendix~\ref{sec:proof} for proof of eqs.~\eqref{L_formula_1}
and~\eqref{L_formula_2}.

\paragraph{\underline{Step 1}}

We first consider the product of
the $\mathcal{Q}$ operator~\eqref{Q_operator} and the
transfer matrix $t_1(\lambda)$.
As we know from the commutation relations~\eqref{commutation_phi} of
the lattice
free field that
a difference field $\dphi_n^{(a)}$ does not commute only with nearest neighbor
$\dphi_{n\pm 1}^{(a)}$,
we get
\begin{align}
  \mathcal{Q}(\lambda) \cdot t_1(\lambda)
  & =
  \Tr
  \prod_n^\curvearrowleft f(\lambda) \, g(\lambda) \,
  \mathcal{L}_n(\lambda ; \phi) \,
  \mathbf{C}_{n+1} \, \mathbf{X}(\lambda)
  \nonumber
  \\
  & =
  \Tr
  \prod_n^\curvearrowleft
  f(\lambda) \, g(\lambda) \,
  \mathbf{M}^{-1} \,
  \mathbf{C}_{n+1}^{~-1} \, \mathcal{L}_n(\lambda ; \phi) \,
  \mathbf{C}_{n+1} \, \mathbf{X}(\lambda) \, \mathbf{C}_n \,
  \mathbf{M} .
  \label{Q_t_step_1}
\end{align}
Here a  matrix $\mathbf{M}$ denotes a gauge-transformation.
We substitute
\begin{equation*}
  \mathbf{M}
  =
  \begin{pmatrix}
    1 & & \\
    1 & 1 & \\
    1 & & 1
  \end{pmatrix} ,
\end{equation*}
into eq.~\eqref{Q_t_step_1}, and compute each element directly.
We see that both   (2,1) and (3,1) components
vanishes
due to
eqs.~\eqref{L_formula_1}.
Thus we  find that a trace of  the product of $3\times 3$ matrix
reduces to a sum of the product of (1,1) component and a trace of the
product of
remaining $2\times 2$ matrices;
\begin{equation}
  \label{Q_t_step_2}
  \mathcal{Q}(\lambda) \cdot t_1(\lambda)
  =
  \prod_n^\curvearrowleft
  f(\lambda) \, g(\lambda) \,
  \mathrm{e}^{-\lambda - \frac{1}{3} \mathrm{i} \gamma} \cdot
  \mathcal{L}_n(\lambda + \frac{2}{3} \, \mathrm{i} \, \gamma ; \phi)
  +
  \Tr
  \prod_n^\curvearrowleft
  f(\lambda) \, g(\lambda)  \,
  \mathbf{A}_n(\lambda) ,
\end{equation}
where
\begin{equation*}
  \mathbf{A}_n(\lambda)
  =
  \begin{pmatrix}
    -\mathcal{L}_n^{(1)}(\lambda ; \phi) \cdot \mathrm{e}^{-\lambda}
    +\mathcal{L}_n^{(2)}(\lambda ; \phi) \cdot \mathrm{e}^{\lambda}
    &
    -\mathcal{L}_n^{(1)}(\lambda ; \phi)
    +\mathcal{L}_n^{(2)}(\lambda ; \phi) \cdot \mathrm{e}^{-\lambda}
    \\[2mm]
    -\mathcal{L}_n^{(1)}(\lambda ; \phi) \cdot \mathrm{e}^{-\lambda}
    +\mathcal{L}_n^{(3)}(\lambda ; \phi)
    &
    -\mathcal{L}_n^{(1)}(\lambda ; \phi)
    +\mathcal{L}_n^{(3)}(\lambda ; \phi) \cdot \mathrm{e}^{\lambda}
  \end{pmatrix}
  \cdot
  \begin{pmatrix}
    c^{(2)}_n & \\[2mm]
    & c_n^{(3)}
  \end{pmatrix} .
\end{equation*}

We now rewrite above  $2\times 2$ matrix $\mathbf{A}_n(\lambda)$.
Using eqs.~\eqref{L_formula_2},
we have
\begin{align*}
  \mathbf{A}_n(\lambda)
  &=
  \mathrm{e}^{-\frac{1}{3}\mathrm{i}\gamma} \,
  (\mathrm{e}^{3 \lambda} - 1) \,
  \begin{pmatrix}
    c_n^{(1)} \mathrm{e}^{-\lambda} + c_n^{(3)}
    &
    - c_n^{(2)} \mathrm{e}^{-\lambda} 
    \\[2mm]
    c_n^{(1)} \mathrm{e}^{-\lambda}
    &
    c_n^{(1)}
  \end{pmatrix}
  \cdot
  \mathcal{L}_n(\lambda - \frac{2}{3} \, \mathrm{i} \, \gamma; \phi)
  \cdot
  \begin{pmatrix}
    c_n^{(2)} & \\[2mm]
    & c_n^{(3)}
  \end{pmatrix}
  \\
  & =
  \mathrm{e}^{-\lambda} 
  (\mathrm{e}^{3 \lambda} - 1) \,
  \begin{pmatrix}
    \bigl( c_n^{(3)} \bigr)^{-1}
    +
    \bigl( c_n^{(1)} \bigr)^{-1}
    \cdot \mathrm{e}^{\lambda - \frac{2}{3}\mathrm{i} \gamma}
    &
    - \bigl( c_n^{(1)} \bigr)^{-1}
    \\[2mm]
    \bigl( c_n^{(3)} \bigr)^{-1} 
    &
    \bigl( c_n^{(2)} \bigr)^{-1} \cdot
    \mathrm{e}^{\lambda - \frac{2}{3} \mathrm{i} \gamma}
  \end{pmatrix}
  \\
  & \qquad
  \times
  \begin{pmatrix}
    \mathcal{L}_n^{(3)}(\lambda - \frac{2}{3} \, \mathrm{i} \, \gamma; \phi)
    & \\[2mm]
    &
    \mathcal{L}_n^{(1)}(\lambda - \frac{2}{3} \, \mathrm{i} \, \gamma; \phi)
  \end{pmatrix} .
\end{align*}
In the last equality, we have used the commutation
relations~\eqref{commutation_phi} and a condition
$c_n^{(1)} \, c_n^{(2)} \, c_n^{(3)}
= \mathrm{e}^{\frac{1}{3} \mathrm{i} \gamma}$.

As a result we get
\begin{multline}
  \label{result_fundamental}
  \mathcal{Q}(\lambda) \cdot t_1(\lambda)
  -
  \prod_n^\curvearrowleft
  f(\lambda) \, g(\lambda) \,
  \mathrm{e}^{-\lambda - \frac{1}{3} \mathrm{i} \gamma} \cdot
  \mathcal{L}_n(\lambda + \frac{2}{3} \, \mathrm{i} \, \gamma ; \phi)
  \\
  =
  \Tr
  \prod_n^\curvearrowleft
  f(\lambda) \, g(\lambda)  \,
  \mathrm{e}^{-\lambda} 
  (\mathrm{e}^{3 \lambda} - 1) \,
  \begin{pmatrix}
    \bigl( c_n^{(3)} \bigr)^{-1}
    +
    \bigl( c_n^{(1)} \bigr)^{-1}
    \cdot \mathrm{e}^{\lambda - \frac{2}{3}\mathrm{i} \gamma}
    &
    - \bigl( c_n^{(1)} \bigr)^{-1}
    \\[2mm]
    \bigl( c_n^{(3)} \bigr)^{-1} 
    &
    \bigl( c_n^{(2)} \bigr)^{-1} \cdot
    \mathrm{e}^{\lambda - \frac{2}{3} \mathrm{i} \gamma}
  \end{pmatrix}
  \\
  \qquad
  \times
  \begin{pmatrix}
    \mathcal{L}_n^{(3)}(\lambda - \frac{2}{3} \, \mathrm{i} \, \gamma; \phi)
    & \\[2mm]
    &
    \mathcal{L}_n^{(1)}(\lambda - \frac{2}{3} \, \mathrm{i} \, \gamma; \phi)
  \end{pmatrix} .
\end{multline}
We keep aside this identity for a while.

\paragraph{\underline{Step 2}}


In the same manner with Step~1, we study a product of the
$\mathcal{Q}$ operator
and the transfer matrix $t_2(\lambda)$.
We have
\begin{align*}
  & \mathcal{Q}(\lambda) \cdot t_2(\lambda)
  \\
  & =
  \Tr
  \prod_n^\curvearrowleft f(\lambda) \, \overline{g}(\lambda) \,
  \mathcal{L}_n(\lambda ; \phi) \,
  \mathbf{C}_{n+1}^{~-1} \, \overline{\mathbf{X}}(\lambda)
  \\
  & =
  \Tr
  \prod_n^\curvearrowleft
  f(\lambda) \, \overline{g}(\lambda) \,
  \mathbf{C}_{n+1} \, \mathcal{L}_n(\lambda ; \phi) \,
  \mathbf{C}_{n+1}^{~-1}  \,
  \overline{\mathbf{X}}(\lambda) \, \mathbf{C}_n^{~-1} \,
  \\[2mm]
  & =
  \Tr
  \prod_n^\curvearrowleft
  f(\lambda) \, \overline{g}(\lambda) \,
  \mathbf{M} \cdot
  \\
  & \qquad \times
  \begin{pmatrix}
    ( c_n^{(1)} )^{-1} \mathrm{e}^\lambda \cdot
    \mathcal{L}_n^{(3)}(\lambda ; \phi)
    &
    - ( c_n^{(2)} )^{-1} \cdot
    \mathcal{L}_n^{(2)}(\lambda ; \phi)
    &
    0
    \\[2mm]
    0
    &
    ( c_n^{(2)} )^{-1} \mathrm{e}^\lambda \cdot
    \mathcal{L}_n^{(1)}(\lambda ; \phi)
    &
    - ( c_n^{(3)} )^{-1} \cdot
    \mathcal{L}_n^{(3)}(\lambda ; \phi)
    \\[2mm]
    - ( c_n^{(1)} )^{-1} \cdot
    \mathcal{L}_n^{(1)}(\lambda ; \phi)
    &
    0
    &
    ( c_n^{(3)} )^{-1} \mathrm{e}^\lambda \cdot
    \mathcal{L}_n^{(2)}(\lambda ; \phi)
  \end{pmatrix}
  \cdot
  \mathbf{M}^{-1} .
\end{align*}
In the last expression, we have introduced matrix $\mathbf{M}$ for
gauge-transformation.
When we substitute
\begin{equation*}
  \mathbf{M}
  =
  \begin{pmatrix}
    1 & & \\
    & 1 & \\
    1 & 1 & 1
  \end{pmatrix} ,
\end{equation*}
we see that  both  (3,1) and (3,2) elements  vanish
due to eqs.~\eqref{L_formula_2}, and that a trace of the product of
$3\times 3$ matrices  can be rewritten as a sum of the product of
(3,3)-element and a trace of the product of $2\times 2$ matrices;
\begin{equation}
  \mathcal{Q}(\lambda) \cdot t_2(\lambda)
  =
  \prod_n^\curvearrowleft
  f(\lambda) \, \overline{g}(\lambda) \,
  \mathrm{e}^{-\frac{1}{3} \mathrm{i} \gamma} \,
  ( \mathrm{e}^{3 \lambda} - 1) \cdot
  \mathcal{L}_n(\lambda - \frac{2}{3} \,  \mathrm{i} \, \gamma; \phi)
  +
  \Tr
  \prod_n^\curvearrowleft
  f(\lambda) \, \overline{g}(\lambda) \,
  \mathbf{B}_n(\lambda) ,
\end{equation}
where $2\times 2$ matrix $\mathbf{B}_n(\lambda)$ is given by
\begin{equation*}
  \mathbf{B}_n(\lambda)
  =
  \begin{pmatrix}
    (c_n^{(1)} )^{-1} \mathrm{e}^\lambda
    \cdot \mathcal{L}_n^{(3)}(\lambda ; \phi)
    &
    - (c_n^{(2)})^{-1}
    \cdot \mathcal{L}_n^{(2)}(\lambda ; \phi)
    \\[2mm]
    (c_n^{(3)})^{-1}
    \cdot \mathcal{L}_n^{(3)}(\lambda ; \phi)
    &
    (c_n^{(2)})^{-1} \mathrm{e}^\lambda
    \cdot \mathcal{L}_n^{(1)}(\lambda ; \phi)
    +
    (c_n^{(3)})^{-1} 
    \cdot \mathcal{L}_n^{(3)}(\lambda ; \phi)
  \end{pmatrix} .
\end{equation*}
We see  that  by applying eqs.~\eqref{L_formula_2} to (2,2)-element of
above  matrix $\mathbf{B}_n(\lambda)$ we have
\begin{multline*}
  \Tr \prod_n^\curvearrowleft
  f(\lambda) \, \overline{g}(\lambda) \,
  \mathbf{B}_n(\lambda)
  \\
  =
  \Tr \prod_n^\curvearrowleft
  f(\lambda) \, \overline{g}(\lambda) \,
  \mathbf{M} \cdot
  \begin{pmatrix}
    (c_n^{(1)} )^{-1} \mathrm{e}^\lambda
    \cdot \mathcal{L}_n^{(3)}(\lambda ; \phi)
    &
    - (c_n^{(2)})^{-1}
    \cdot \mathcal{L}_n^{(2)}(\lambda ; \phi)
    \\[2mm]
    (c_n^{(3)})^{-1}
    \cdot \mathcal{L}_n^{(3)}(\lambda ; \phi)
    &
    \Bigl(
    (c_n^{(3)})^{-1} \mathrm{e}^\lambda
    + (c_n^{(2)})^{-1}
    \Bigr) \cdot
    \mathcal{L}_n^{(2)}(\lambda; \phi)
  \end{pmatrix}
  \cdot
  \mathbf{M}^{-1} ,
\end{multline*}
where  the invertible matrix $\mathbf{M}$ is included again for
gauge-transformation.
Substituting
\begin{equation*}
  \mathbf{M}
  =
  \begin{pmatrix}
    1 & 1 \\
    & 1
  \end{pmatrix} ,
\end{equation*}
and erasing $\mathcal{L}_n^{(2)}(\lambda; \phi)$
using  eqs.~\eqref{L_formula_2}, we get
\begin{multline*}
  \Tr \prod_n^\curvearrowleft
  f(\lambda) \, \overline{g}(\lambda) \,
  \mathbf{B}_n(\lambda)
  \\
  =
  \Tr \prod_n^\curvearrowleft
  f(\lambda) \, \overline{g}(\lambda) \,
  \begin{pmatrix}
    (c_n^{(1)})^{-1} \mathrm{e}^\lambda + (c_n^{(3)})^{-1}
    &
    - (c_n^{(1)})^{-1}
    \\[2mm]
    (c_n^{(3)})^{-1}
    &
    (c_n^{(2)})^{-1} \mathrm{e}^\lambda
  \end{pmatrix}
  \cdot
  \begin{pmatrix}
    \mathcal{L}_n^{(3)}(\lambda ; \phi) &
    \\[2mm]
    & \mathcal{L}_n^{(1)}(\lambda ; \phi)
  \end{pmatrix} .
\end{multline*}

As a result we obtain an identity,
\begin{multline}
  \label{result_adjoint}
  \mathcal{Q}(\lambda) \cdot t_2(\lambda)
  -
  \prod_n^\curvearrowleft
  f(\lambda) \, \overline{g}(\lambda) \,
  \mathrm{e}^{-\frac{1}{3} \mathrm{i} \gamma} \,
  ( \mathrm{e}^{3 \lambda} - 1) \cdot
  \mathcal{L}_n(\lambda - \frac{2}{3} \, \mathrm{i} \, \gamma; \phi)
  \\
  =
  \Tr \prod_n^\curvearrowleft
  f(\lambda) \, \overline{g}(\lambda) \,
  \begin{pmatrix}
    (c_n^{(1)})^{-1} \mathrm{e}^\lambda + (c_n^{(3)})^{-1}
    &
    - (c_n^{(1)})^{-1}
    \\[2mm]
    (c_n^{(3)})^{-1}
    &
    (c_n^{(2)})^{-1} \mathrm{e}^\lambda
  \end{pmatrix}
  \cdot
  \begin{pmatrix}
    \mathcal{L}_n^{(3)}(\lambda ; \phi) &
    \\[2mm]
    & \mathcal{L}_n^{(1)}(\lambda ; \phi)
  \end{pmatrix} .
\end{multline}

\paragraph{\underline{Step 3}}

By  comparing eq.~\eqref{result_fundamental} and
eq.~\eqref{result_adjoint},
we find  that  $2 \times 2$ matrices in the right hand sides
have same form with each
other.
By erasing these  $2 \times 2$ matrices, we obtain an identity for
operators $t_1(\lambda)$, $t_2(\lambda)$, and $\mathcal{Q}(\lambda)$;
\begin{multline*}
  \mathcal{Q}(\lambda) \cdot t_1(\lambda)
  -
  \left(
    \prod_n^\curvearrowleft 
    \frac{f(\lambda)}
    {f(\lambda+\frac{2}{3} \, \mathrm{i} \, \gamma)} \,
    g(\lambda) \,
    \mathrm{e}^{-\lambda - \frac{1}{3} \mathrm{i} \gamma}
  \right) \cdot
  \mathcal{Q}(\lambda + \frac{2}{3} \, \mathrm{i} \, \gamma)
  \\
  =
  \left(
    \prod_n^\curvearrowleft
    \frac{f(\lambda) \, g(\lambda)}
    { f(\lambda - \frac{2}{3} \, \mathrm{i} \, \gamma) \,
      \overline{g}(\lambda - \frac{2}{3} \, \mathrm{i} \, \gamma)}
    \,
    \mathrm{e}^{- \lambda} \,
    (\mathrm{e}^{3 \lambda } - 1)
  \right) \,
  \mathcal{Q}(\lambda - \frac{2}{3} \, \mathrm{i} \, \gamma) \cdot
  t_2(\lambda - \frac{2}{3} \, \mathrm{i} \, \gamma)
  \\
  -
  \left(
    \prod_n^\curvearrowleft
    \frac{f(\lambda)}{f(\lambda - \frac{4}{3} \, \mathrm{i} \,
      \gamma)}
    \, g(\lambda) \,
    \mathrm{e}^{-\lambda - \frac{1}{3} \mathrm{i} \gamma}
    (\mathrm{e}^{3 \lambda} - 1) \,
    (\mathrm{e}^{3 \lambda - 2 \mathrm{i} \gamma} - 1)
  \right) \,
  \mathcal{Q}(\lambda - \frac{4}{3} \, \mathrm{i} \, \gamma) .
\end{multline*}
After setting normalization functions as
\begin{align*}
  f(\lambda) & = 1 , \\[2mm]
  g(\lambda) & = \mathrm{e}^{\lambda + \frac{1}{3} \mathrm{i} \gamma}
  ,
  \\[2mm]
  \overline{g}(\lambda)
  & =
  \mathrm{e}^{\frac{1}{3} \mathrm{i} \gamma} \,
  \bigl(
  \mathrm{e}^{3 \lambda + 2 \mathrm{i} \gamma} - 1
  \bigr) ,
\end{align*}
we see
that the operator $\mathcal{Q}(\lambda)$ solves the
difference equation,
\begin{multline}
  \mathcal{Q}(\lambda + 2 \, \mathrm{i} \, \gamma)
  -
  t_1(\lambda+ \frac{4}{3} \, \mathrm{i} \, \gamma)
  \cdot \mathcal{Q}(\lambda + \frac{4}{3} \, \mathrm{i} \, \gamma)
  \\
  +
  t_2(\lambda + \frac{2}{3} \, \mathrm{i} \, \gamma)
  \cdot
  \mathcal{Q}(\lambda + \frac{2}{3} \, \mathrm{i} \, \gamma)
  -
  \Delta(\lambda + \frac{2}{3} \, \mathrm{i} \, \gamma) \cdot
  \mathcal{Q}(\lambda)
  = 0 ,
  \label{Baxter_1}
\end{multline}
where the function $\Delta(\lambda)$ is defined by
\begin{equation*}
  \Delta(\lambda)
  =
  \bigl(
  (  \mathrm{e}^{3  \lambda } - 1 ) \,
  (  \mathrm{e}^{3  \lambda + 2 \mathrm{i} \gamma } - 1 ) \,
  \bigr)^{L} .
\end{equation*}
As we know from eq.~\eqref{Q_commute_t}
that the  transfer matrices $t_{1,2}(\lambda)$
and the $\mathcal{Q}$ operator
commute each other,
the operator valued
difference equation~\eqref{Baxter_1} can be regarded as the
third order difference equation in usual sense,
once we apply both hand sides to
simultaneous eigenfunction.

We expect
in general the third order difference equation~\eqref{Baxter_1}
has three linearly
independent solutions, which we write
$\mathcal{Q}_{a}(\lambda)$ for $a=+, 0, -$.
We then introduce the function
$\mathcal{P}(\lambda)$ as the quantum Wronskian of solutions;
\begin{equation}
  \label{P_Wronskian}
  \mathcal{P}_{ab}(\lambda)
  =
  \begin{vmatrix}
    \mathcal{Q}_a(\lambda)
    &
    \mathcal{Q}_a(\lambda - \frac{2}{3} \, \mathrm{i} \, \gamma)
    \\[2mm]
    \mathcal{Q}_b(\lambda)
    &
    \mathcal{Q}_b(\lambda - \frac{2}{3} \, \mathrm{i} \, \gamma)
  \end{vmatrix} ,
\end{equation}
for $a,b=0,\pm$
(we thus have three independent functions $\mathcal{P}_{ab}(\lambda)$).
It is straightforward to see that the function
$\mathcal{P}_{ab}(\lambda)$
satisfies the third order difference equation,
\begin{multline}
  \mathcal{P}(\lambda + 2 \, \mathrm{i} \, \gamma)
  -
  t_2(\lambda + \frac{2}{3} \, \mathrm{i} \, \gamma )
  \cdot
  \mathcal{P}(\lambda + \frac{4}{3} \, \mathrm{i} \, \gamma)
  \\
  +
  \Delta(\lambda + \frac{2}{3} \, \mathrm{i} \, \gamma) \,
  t_1(\lambda + \frac{2}{3} \, \mathrm{i} \, \gamma) \cdot
  \mathcal{P}(\lambda + \frac{2}{3} \, \mathrm{i} \, \gamma)
  -
  \Delta(\lambda) \,
  \Delta(\lambda + \frac{2}{3} \, \mathrm{i} \, \gamma)    \cdot
  \mathcal{P}(\lambda)
  = 0 .
  \label{Baxter_2}
\end{multline}


{}From two difference equations~\eqref{Baxter_1} and~\eqref{Baxter_2},
we get the Baxter equation for the discrete Boussinesq equation;
\begin{subequations}
  \label{diagonalize_t}
  \begin{align}
    t_1(\lambda)
    & =
    \Delta(\lambda - \frac{2}{3} \, \mathrm{i} \, \gamma) \,
    \frac{\mathcal{P}(\lambda - \frac{2}{3} \, \mathrm{i} \, \gamma)}
    {\mathcal{P}(\lambda)}
    +
    \frac{\mathcal{P}(\lambda+ \frac{2}{3} \, \mathrm{i} \, \gamma)}
    {\mathcal{P}(\lambda)}
    \cdot
    \frac{\mathcal{Q}(\lambda - \frac{2}{3} \, \mathrm{i} \, \gamma)}
    {\mathcal{Q}(\lambda)}
    +
    \frac{\mathcal{Q}(\lambda + \frac{2}{3} \, \mathrm{i} \, \gamma)}
    {\mathcal{Q}(\lambda)} ,
    \\[2mm]
    t_2(\lambda)
    & =
    \frac{\mathcal{P}(\lambda + \frac{4}{3} \, \mathrm{i} \, \gamma)}
    {\mathcal{P}(\lambda + \frac{2}{3} \, \mathrm{i} \, \gamma)}
    +
    \Delta(\lambda) \,
    \frac{\mathcal{P}(\lambda)}
    {\mathcal{P}(\lambda  + \frac{2}{3} \, \mathrm{i} \, \gamma )}
    \cdot
    \frac{\mathcal{Q}(\lambda + \frac{2}{3} \, \mathrm{i} \, \gamma)}
    {\mathcal{Q}(\lambda)}
    +
    \Delta(\lambda) \,
    \frac{\mathcal{Q}(\lambda - \frac{2}{3} \, \mathrm{i} \, \gamma)}
    {\mathcal{Q}(\lambda)} .
  \end{align}
\end{subequations}
Once we have a set of functional equations~\eqref{diagonalize_t}  we
can follow a strategy in Ref.~\citen{KuliResh86} to obtain the
spectrum of these transfer matrices;
we suppose
\begin{align}
  \mathcal{Q}(\lambda)
  & =
  \prod_n \sh(\lambda - \lambda_n^{(1)}) ,
  &
  \mathcal{P}(\lambda)
  & =
  \prod_n \sh(\lambda - \lambda_n^{(2)}) ,
\end{align}
and  substitute them into eqs.~\eqref{diagonalize_t}.
{}From the condition of analyticity of the transfer matrices
$t_a(\lambda)$
we get the nested Bethe ansatz equations.
The precise analysis of these equations
is for future studies.
As a consequence,
we have diagonalized   the transfer matrices
$t_{1,2}(\lambda)$
by use  of the auxiliary functions
$\mathcal{Q}(\lambda)$ and $\mathcal{P}(\lambda)$.

\section{Duality}
\label{sec:duality}

We have seen that the $\mathcal{Q}$ operator~\eqref{Q_operator}
satisfies
the third order
difference equation~\eqref{Baxter_1}
whose coefficients are given by two transfer matrices
$t_1(\lambda)$ and $t_2(\lambda)$ for the quantum discrete Boussinesq
equation.
As was noticed in Appendix~\ref{sec:dilog} the $\mathcal{Q}$
operator
has a property of duality which follows from
eq.~\eqref{dual_intertwine} and eq.~\eqref{duality_Phi};
\begin{align}
  \label{duality_transform}
  \gamma
  & \to
  \frac{\pi^2}{\gamma} ,
  &
  \phi_n^{(a)}
  & \to
  \frac{\pi}{\gamma} \, \phi_n^{(a)} ,
  &
  \lambda
  & \to
  \frac{\pi}{\gamma} \, \lambda  .
\end{align}
Above duality  implies the existence of the dual Baxter equation;
we have the dual third order difference equation,
\begin{multline}
  \label{dual_Baxter_1}
  \mathcal{Q}(\lambda + 2 \, \mathrm{i} \, \pi)
  -
  {u}_1(\lambda + \frac{4}{3} \, \mathrm{i} \, \pi) \cdot
  \mathcal{Q}(\lambda + \frac{4}{3} \, \mathrm{i} \, \pi)
  \\
  +
  {u}_2(\lambda + \frac{2}{3} \, \mathrm{i} \, \pi)
  \cdot
  \mathcal{Q}(\lambda + \frac{2}{3} \, \mathrm{i} \, \pi)
  -
  \Tilde{\Delta}(\lambda + \frac{2}{3} \, \mathrm{i} \, \pi) \cdot
  \mathcal{Q}(\lambda)
  = 0 ,
\end{multline}
where
\begin{equation*}
  \Tilde{\Delta}(\lambda)
  =
  \Bigl(
  (  \mathrm{e}^{3 \frac{\pi}{\gamma}  \lambda } - 1 ) \,
  (  \mathrm{e}^{
    \frac{\pi}{\gamma} ( 3  \lambda + 2 \mathrm{i} \pi)
    } - 1 ) \,
  \Bigr)^{L} .
\end{equation*}
The operator $\mathcal{Q}(\lambda)$ is as same with
eq.~\eqref{Q_operator}, and
the transfer matrix ${u}_{a}(\lambda)$ is defined by
\begin{align}
  {u}_1(\lambda)
  & =
  \Tr
  {\mathbf{U}}(\lambda) ,
  &
  {\mathbf{U}}(\lambda)
  & =
  \prod_n^\curvearrowleft
  {\mathbf{M}}_n(\lambda)  ,
  \\[2mm]
  {u}_2(\lambda)
  & =
  \Tr
  \overline{\mathbf{U}}(\lambda) ,
  &
  \overline{\mathbf{U}}(\lambda)
  & =
  \prod_n^\curvearrowleft
  \overline{\mathbf{M}}_n(\lambda)  ,
\end{align}
with the dual Lax matrix $\mathbf{M}(\lambda)$ and
$\overline{\mathbf{M}}(\lambda)$,
\begin{align*}
  \mathbf{M}_n(\lambda)
  & =
  h(\lambda) \,
  \begin{pmatrix}
    \mathrm{e}^{
      \frac{\pi}{\gamma}
      \bigl(
      \frac{1}{3}\dphi_n^{(1)} - \frac{1}{3} \dphi_n^{(2)}
      \bigr)
      }
    &&\\
    & \mathrm{e}^{
      \frac{\pi}{\gamma}
      \bigl(
      \frac{1}{3}\dphi_n^{(1)} + \frac{2}{3} \dphi_n^{(2)}
      \bigr)
      }
    &\\
    && \mathrm{e}^{
      \frac{\pi}{\gamma}
      \bigl(
      -\frac{2}{3}\dphi_n^{(1)} - \frac{1}{3} \dphi_n^{(2)}
      \bigr)
      }
  \end{pmatrix}
  \cdot
  \begin{pmatrix}
    \mathrm{e}^{\frac{\pi}{\gamma} \lambda}
    & \mathrm{e}^{- \frac{\pi}{\gamma} \lambda} & 1 \\[2mm]
    1 & \mathrm{e}^{\frac{\pi}{\gamma} \lambda}
    & \mathrm{e}^{-\frac{\pi}{\gamma} \lambda} \\[2mm]
    \mathrm{e}^{-\frac{\pi}{\gamma} \lambda} & 1  &
    \mathrm{e}^{\frac{\pi}{\gamma} \lambda }
  \end{pmatrix} ,
  \\[2mm]
  \overline{\mathbf{M}}_n(\lambda)
  & =
  \overline{h}(\lambda) \,
  \begin{pmatrix}
    \mathrm{e}^{
      \frac{\pi}{\gamma}
      \bigl(
        -\frac{1}{3}\dphi_n^{(1)} + \frac{1}{3} \dphi_n^{(2)}
      \bigr)
      }
    &&\\
    & \mathrm{e}^{
      \frac{\pi}{\gamma}
      \bigl(
        -\frac{1}{3}\dphi_n^{(1)} - \frac{2}{3} \dphi_n^{(2)}
      \bigr)
      }
    &\\
    && \mathrm{e}^{
      \frac{\pi}{\gamma}
      \bigl(
        \frac{2}{3}\dphi_n^{(1)} + \frac{1}{3} \dphi_n^{(2)}
      \bigr)
      }
  \end{pmatrix}
  \cdot
  \begin{pmatrix}
    \mathrm{e}^{\frac{\pi}{\gamma} \lambda} & -1 & 0 \\[2mm]
    0 &\mathrm{e}^{\frac{\pi}{\gamma} \lambda} & -1  \\[2mm]
    -1 & 0 & \mathrm{e}^{\frac{\pi}{\gamma} \lambda }
  \end{pmatrix} .
\end{align*}

To get the dual Baxter equation,
another operator $\mathcal{P}$ should be
slightly different from eq.~\eqref{P_Wronskian},
\begin{equation}
  \Tilde{\mathcal{P}}_{ab}(\lambda)
  =
  \begin{vmatrix}
    \mathcal{Q}_a(\lambda)
    &
    \mathcal{Q}_a(\lambda - \frac{2}{3} \, \mathrm{i} \, \pi)
    \\[2mm]
    \mathcal{Q}_b(\lambda)
    &
    \mathcal{Q}_b(\lambda - \frac{2}{3} \, \mathrm{i} \, \pi)
  \end{vmatrix}  ,
\end{equation}
where $\mathcal{Q}_a(\lambda)$ denotes three linearly independent
solutions of the difference equation~\eqref{dual_Baxter_1}.
This solves  the third order difference equation,
\begin{multline}
  \Tilde{\mathcal{P}}(\lambda + 2 \, \mathrm{i} \, \pi)
  -
  u_2(\lambda + \frac{2}{3} \, \mathrm{i} \, \pi )
  \cdot
  \Tilde{\mathcal{P}}(\lambda + \frac{4}{3} \, \mathrm{i} \, \pi)
  \\
  +
  \Tilde{\Delta}(\lambda + \frac{2}{3} \, \mathrm{i} \, \pi) \,
  u_1(\lambda + \frac{2}{3} \, \mathrm{i} \, \pi) \cdot
  \Tilde{\mathcal{P}}(\lambda + \frac{2}{3} \, \mathrm{i} \, \pi)
  -
  \Tilde{\Delta}(\lambda) \,
  \Tilde{\Delta}(\lambda + \frac{2}{3} \, \mathrm{i} \, \pi)    \cdot
  \Tilde{\mathcal{P}}(\lambda)
  = 0 .
  \label{dual_Baxter_2}
\end{multline}
As a result,
we obtain the dual Baxter equations,
\begin{subequations}
  \begin{align}
    u_1(\lambda)
    & =
    \Tilde{\Delta}(\lambda - \frac{2}{3} \, \mathrm{i} \, \pi) \,
    \frac{
      \Tilde{\mathcal{P}}(\lambda - \frac{2}{3} \, \mathrm{i} \, \pi)}
    {\Tilde{\mathcal{P}}(\lambda)}
    +
    \frac{
      \Tilde{\mathcal{P}}(\lambda+ \frac{2}{3} \, \mathrm{i} \, \pi)}
    {\Tilde{\mathcal{P}}(\lambda)}
    \cdot
    \frac{\mathcal{Q}(\lambda - \frac{2}{3} \, \mathrm{i} \, \pi)}
    {\mathcal{Q}(\lambda)}
    +
    \frac{\mathcal{Q}(\lambda + \frac{2}{3} \, \mathrm{i} \, \pi)}
    {\mathcal{Q}(\lambda)} ,
    \\[2mm]
    u_2(\lambda)
    & =
    \frac{
      \Tilde{\mathcal{P}}(\lambda + \frac{4}{3} \, \mathrm{i} \, \pi)}
    {\Tilde{\mathcal{P}}(\lambda + \frac{2}{3} \, \mathrm{i} \, \pi)}
    +
    \Tilde{\Delta}(\lambda) \,
    \frac{\Tilde{\mathcal{P}}(\lambda)}
    {\Tilde{\mathcal{P}}(\lambda  + \frac{2}{3} \, \mathrm{i} \, \pi )}
    \cdot
    \frac{\mathcal{Q}(\lambda + \frac{2}{3} \, \mathrm{i} \, \pi)}
    {\mathcal{Q}(\lambda)}
    +
    \Tilde{\Delta}(\lambda) \,
    \frac{\mathcal{Q}(\lambda - \frac{2}{3} \, \mathrm{i} \, \pi)}
    {\mathcal{Q}(\lambda)} ,
  \end{align}
\end{subequations}
from which we get the  nested Bethe ansatz equations only from the
analytic property of the transfer matrices $u_a(\lambda)$.

We remark that this duality of the Baxter equation
also appeared in studies of the discrete KdV
equation~\cite{FSmir00a}, where given was the interpretation from the
viewpoint of the algebraic geometry.

%
%
\section{Continuum Limit}
\label{sec:continue}

We briefly study the continuum limit of the quantum discrete
Boussinesq equation.
In a continuum limit,
the lattice free field $\phi_n^{(a)}$ is replaced by $\phi^{(a)}(x)$,
and
the commutation relations~\eqref{commutation_phi} then  reduce into
\begin{align}
  [\phi^{(a)}(x) ~,~ \phi^{(b)}(y)]
  =
  \mathbf{C}_{ab} \, \mathrm{i} \, \gamma \, \sgn(x-y) ,
\end{align}
where $\mathbf{C}$ is the Cartan matrix of $s\ell_3$,
\begin{equation*}
  \mathbf{C}
  =
  \begin{pmatrix}
    2& -1 \\
    -1 & 2
  \end{pmatrix} .
\end{equation*}
The quasi-periodic boundary condition~\eqref{lattice_boundary}
denotes
\begin{equation*}
  \phi^{(a)}(x+L)
  =
  \phi^{(a)}(x)
  + L \cdot \mathrm{i} \, P^{(a)} ,
\end{equation*}
which indicates the mode expansion of the free fields
as~\cite{FateLuky92b}
\begin{equation*}
  \phi^{(b)}(x)
  =
  \mathrm{i} \, Q^{(b)} + \mathrm{i} \, P^{(b)} \, x
  +
  \sum_{n \neq 0}
  \frac{a_{-n}^{(b)}}{n} \, 
  \mathrm{e}^{\mathrm{i} \frac{2 \pi}{L} n x}  .
\end{equation*}

We can take  the continuum limit of the transfer matrices
$t_{1,2}(\lambda)$ from  eq.~\eqref{rewrite_t1} and
eq.~\eqref{rewrite_t2}.
After a gauge-transformation, we get
\begin{align}
  & \bigl(
  g(\lambda) \,
  \mathrm{e}^{\lambda + \frac{1}{3} \mathrm{i} \gamma}
  \bigr)^{-L} \cdot
  t_1(\lambda)
  \nonumber \\
  & \quad
  \to
  \Tr
  \biggl[
  \mathrm{e}^{\mathrm{i} L
    \bigl(
    P^{(1)} \mathbf{h}_1
    -
    P^{(2)} \mathbf{h}_3
    \bigr)
    }
  \,
  \mathscr{P}
  \exp
  \left(
    \int_0^L
    \mathrm{d} x
    \begin{pmatrix}
      0 &
      \mathrm{e}^{-2 \lambda - \frac{1}{6}\mathrm{i}\gamma} \,
      \mathrm{e}^{-\phi^{(1)}(x)}
      &
      \mathrm{e}^{- \lambda + \frac{1}{6}\mathrm{i}\gamma} \,
      \mathrm{e}^{\phi^{(0)}(x)}
      \\
      \mathrm{e}^{- \lambda + \frac{1}{6}\mathrm{i}\gamma} \,
      \mathrm{e}^{\phi^{(1)}(x)}
      & 0
      &
      \mathrm{e}^{-2 \lambda - \frac{1}{6}\mathrm{i}\gamma} \,
      \mathrm{e}^{-\phi^{(2)}(x)}
      \\
      \mathrm{e}^{-2 \lambda - \frac{1}{6}\mathrm{i}\gamma} \,
      \mathrm{e}^{-\phi^{(0)}(x)}
      &
      \mathrm{e}^{- \lambda + \frac{1}{6}\mathrm{i}\gamma} \,
      \mathrm{e}^{\phi^{(2)}(x)}
      & 0
    \end{pmatrix}
  \right)
  \biggr] ,
  \label{continue_t1}
  \\[2mm]
  & \bigl(
  \overline{g}(\lambda) \,
  \mathrm{e}^{\lambda + \frac{1}{3} \mathrm{i} \gamma}
  \bigr)^{-L}
  \cdot
  t_2(\lambda)
  \nonumber \\
  & \quad
  \to
  \Tr
  \biggl[
  \mathrm{e}^{\mathrm{i} L
    \bigl(
    -P^{(1)} \mathbf{h}_1
    + P^{(2)} \mathbf{h}_3
    \bigr)
    }
  \,
  \mathscr{P}
  \exp
  \left(
    -     \mathrm{e}^{-\lambda - \frac{1}{6} \mathrm{i} \gamma}
    \,
    \int_0^L
    \mathrm{d} x \,
    \begin{pmatrix}
      0 & \mathrm{e}^{\phi^{(1)}}(x) & 0 \\
      0 & 0 & \mathrm{e}^{\phi^{(2)}(x)} \\
      \mathrm{e}^{\phi^{(0)}(x)} & 0 & 0
    \end{pmatrix}
  \right)
  \biggr] ,
  \label{continue_t2}
\end{align}
where $\mathscr{P}$ denotes the path operator ordering.
One finds that
the transfer matrix $t_2(\lambda)$ in eq.~\eqref{continue_t2}
essentially coincides with that
was proposed in~\citen{FateLuky92b} as a quantized Drinfeld--Sokolov
reduction.

Correspondingly a difference of the free fields~\eqref{define_dphi}
becomes the
differential of the free field,
\begin{equation*}
  \dphi_n^{(a)} \to - \frac{\partial}{\partial x} \phi^{(a)}(x) ,
\end{equation*}
and  the Baxter $\mathcal{Q}$ operator for the quantum Boussinesq
equation can be given by the path ordering of the operator
$\mathcal{L}_n(\lambda ;\phi)$.
As a result  we recover
the third order difference equation equation~\eqref{Baxter_1}.
At this stage we are not sure how to relate the $\mathcal{Q}$ operator
with the universal $\mathcal{R}$ matrix.

\section{Generalization: $N$-reduced Discrete KP Equation}
\label{sec:kp}

As the  Boussinesq equation~\eqref{usual_Boussinesq} is the
3-reduced KP equation,
we can introduce the quantum
$N$-reduced discrete KP equation which is related with the affine
$\widehat{s\ell}_N$ Toda field theory.
In the classical case,
the Lax matrix for the discrete $N$-reduced KP equation is given
by~\cite{HikamRInou97b,Hikam97b}
\begin{equation}
  \label{general_L_x}
  \mathbf{L}_n^{\mathrm{W}}(x)
  =
  \frac{1}{\sqrt[N]{W_n^{(N)}}}
  \begin{pmatrix}
    x^{N-1} & - x^{N-2} \, W_{n+N-2}^{(2)} &
    x^{N-3} \, W_{n+N-3}^{(3)}
    & \cdots &
    (-1)^{N-2} \, x \, W_{n+1}^{(N-1)}
    & (-1)^{N-1} \, W_n^{(N)}
    \\
    1 & 0 & 0 & \cdots & 0 & 0 \\
    0 & 1 & 0 & \cdots & 0 & 0 \\
    \vdots & 0 & 1 & \ddots & \vdots & \vdots \\
    \vdots & \vdots & \ddots & \ddots & \ddots & \vdots \\
    0 & 0 & \cdots & \cdots & 1 & 0
  \end{pmatrix} ,
\end{equation}
where the generators of the lattice $W_N$ algebra are defined as
\begin{equation}
  \label{Miura_transform}
  \begin{split}
  W_n^{(s)}
  & =
  \frac{1}{
    \displaystyle
    \prod_{k=0}^{s-1}
    \biggl(
    1 +
    \sum_{i=1}^{N-1}
    \exp (\chi_{n+k}^{(i)})
    \biggr)
    }
  \Biggl(
  \sum_{
    1 \leq i_1 < i_2 < \cdots < i_{s-1} \leq N-1
    }
  \exp
  \left(
    \chi_{n+s-1}^{(i_1)}
    +     \chi_{n+s-2}^{(i_2)}
    + \dots +
    \chi_{n+1}^{(i_{s-1})}
  \right)
  \\
  & \qquad \quad
  +
  \sum_{
    1 \leq i_1 < i_2 < \cdots < i_{s} \leq N-1
    }
  \exp
  \left(
    \chi_{n+s-1}^{(i_1)}
    +     \chi_{n+s-2}^{(i_2)}
    + \dots +
    \chi_{n}^{(i_{s})}
  \right)
  \Biggr) ,
  \quad
  \text{for $s=2,3,\dots,N-1$},
  \\[2mm]
  W_n^{(N)}
  & =
  \frac{
    \displaystyle
    \exp
    \left(
      \sum_{i=1}^{N-1}
      \chi_{n+N-i}^{(i)}
    \right)
    }{
    \displaystyle
    \prod_{k=0}^{N-1}
    \left(
      1 +
      \sum_{i=1}^{N-1}
      \exp(\chi_{n+k}^{(i)})
    \right)
    } .
\end{split}
\end{equation}
Here $\chi_n^{(a)}$ is defined in terms of the free fields
$\phi_n^{(a)}$
(for $a=1,2,\dots,N-1$)
as
\begin{equation*}
  \chi_n^{(a)}
  =
  \sum_{b=1}^a
  \dphi_n^{(b)} ,
\end{equation*}
and the quantum algebra for the lattice free fields reads as
\begin{equation}
  \begin{split}
    [\phi_m^{(a)} ~,~ \phi_n^{(a)}]
    &= 2 \, \mathrm{i} \, \gamma  , \quad \text{for $m>n$},
%
    \\
    [\phi_n^{(a)} ~,~ \phi_m^{(a+1)}]
    & =
    \begin{cases}
      \displaystyle
      \mathrm{i} \, \gamma  ,
      & \text{for $n \leq m$},
      \\
      \displaystyle
      - \mathrm{i} \, \gamma ,
      & \text{for $n >m$}.
    \end{cases}
  \end{split}
\end{equation}
We note that
the transformation~\eqref{Miura_transform} is the so-called Miura
transformation.
We set
\begin{equation*}
  \phi_n^{(0)} =
  -\sum_{a=1}^{N-1} \phi_n^{(a)} ,
\end{equation*}
and the screening charges are defined by
(for $a=0, 1, \dots, N-1$)
\begin{align}
  Q^{(a)}
  & =
  \sum_n \mathrm{e}^{\phi_n^{(a)}} ,
  &
  \Tilde{Q}^{(a)}
  & =
  \sum_n \mathrm{e}^{\frac{\pi}{\gamma}\phi_n^{(a)}} .
\end{align}
The sum of the screening charges,
$\sum_{a=0}^{N-1} Q^{(a)}$, corresponds to the Hamiltonian of the
affine Toda field theory.

In terms of the free fields,
the quantum Lax matrix is then defined by~\cite{Hikam97b}
\begin{equation}
  \mathbf{L}_n(\lambda)
  =
  \exp
  \biggl(
  \sum_{a=1}^{N-1}
  \mathbf{h}_a \, \chi_n^{(a)}
  \biggr) \cdot
  \mathbf{X}(\lambda) ,
\end{equation}
where $N\times N$ matrices $\mathbf{h}_a$ and $\mathbf{X}(\lambda)$
are defined by
\begin{align*}
  \mathbf{h}_a
  & =
  \mathbf{E}_{a,a} - \frac{1}{N}  \, \openone ,
  &
  \mathbf{X}(\lambda)
  & =
  \sum_{k=0}^{N-1}
  \mathrm{e}^{- k \lambda} \,
  \mathbf{C}^k ,
  &
  \mathbf{C}
  & =
  \mathbf{E}_{1,N} +
  \sum_{k=1}^{N-1} \mathbf{E}_{k+1,k} .
\end{align*}
The spectral parameter $\lambda$ is related with $x$ in
eq.~\eqref{general_L_x} by
\begin{equation*}
  \mathrm{e}^{- N \lambda} = 1 - x^{-N} .
\end{equation*}
As usual the monodromy and the transfer matrix
are respectively given by
\begin{align}
  \label{NKP_monodromy}
  \mathbf{T}(\lambda)
  & =
  \prod_n^\curvearrowleft
  \mathbf{L}_n(\lambda) ,
  &
  t_1(\lambda)
  & =
  \Tr \mathbf{T}(\lambda)  .
\end{align}
This transfer matrix generates the integrals of motion of the discrete
$N$-reduced  KP equation.

The fundamental $\mathcal{L}$ operator for
$s\ell_N$ case is defined
by~\cite{Hikam97c}
\begin{multline}
  \mathcal{L}_n(\lambda ; \phi)
  =
  \frac{1}{\Phi_\gamma(\lambda + \chi_n^{(N-1)})} \cdot
  \frac{1}{\Phi_\gamma(2 \, \lambda + \chi_n^{(N-2)})} \cdots
  \frac{1}{\Phi_\gamma((N-1) \, \lambda + \chi_n^{(1)})}
  \\
  \times
  \frac{\Theta_\gamma(\chi_n^{(1)})}
  {\Phi_\gamma(\lambda - \chi_n^{(1)})} \cdot
  \frac{\Theta_\gamma(\chi_n^{(2)})}
  {\Phi_\gamma(\lambda - \chi_n^{(2)})} \cdots
  \frac{\Theta_\gamma(\chi_n^{(N-1)})}
  {\Phi_\gamma(\lambda - \chi_n^{(N-1)})}  ,
%
\end{multline}
This operator not only satisfies the Yang--Baxter
equation~\eqref{YBE_for_L}, but
fulfills the intertwining relations for the (dual)
vertex operators~\cite{Hikam98c};
\begin{gather}
  \Bigl(
  \mathrm{e}^{\phi_n^{(a)}}
  + \mathrm{e}^\lambda \, \mathrm{e}^{\phi_{n+1}^{(a)}}
  \Bigr) \cdot
  \mathcal{L}_n(\lambda ;  \phi)
  =
  \mathcal{L}_n(\lambda ;  \phi)
  \cdot
  \Bigl(
  \mathrm{e}^\lambda \, \mathrm{e}^{\phi_n^{(a)}}
  +  \mathrm{e}^{\phi_{n+1}^{(a)}}
  \Bigr) ,
  \\[2mm]
  \Bigl(
  \mathrm{e}^{\frac{\pi}{\gamma} \phi_n^{(a)}}
  + \mathrm{e}^{\frac{\pi}{\gamma} \lambda}  \,
  \mathrm{e}^{\frac{\pi}{\gamma} \phi_{n+1}^{(a)}}
  \Bigr) \cdot
  \mathcal{L}_n(\lambda ;  \phi )
  =
  \mathcal{L}_n(\lambda ;  \phi )
  \cdot
  \Bigl(
  \mathrm{e}^{\frac{\pi}{\gamma} \lambda} \,
  \mathrm{e}^{\frac{\pi}{\gamma} \phi_n^{(a)}}
  +  \mathrm{e}^{\frac{\pi}{\gamma} \phi_{n+1}^{(a)}}
  \Bigr) ,
\end{gather}
for $a=0,1,\dots,N-1$.
We can easily see that the $\mathcal{Q}$ operator
\begin{equation}
  \label{Q_general}
  \mathcal{Q}(\lambda)
  =
  \prod_n^\curvearrowleft
  \mathcal{L}_n(\lambda ; \phi) ,
\end{equation}
commute with the screening charges
\begin{equation}
  [ \mathcal{Q}(\lambda) ~,~ Q^{(a)} ]
  =
  [ \mathcal{Q}(\lambda) ~,~ \Tilde{Q}^{(a)} ]
  = 0  ,
\end{equation}
which proves that the $\mathcal{Q}$ operator generates the local
integrals of motion for the discrete analogue of the affine
$\widehat{s\ell}_N$ Toda field theory.

As we have clarified for $s\ell_3$ case
in preceding sections, the generating function of
the local integrals of motion for the affine Toda field theory becomes
the $\mathcal{Q}$ operator for the quantum Boussinesq equation.
Thus it is natural to conjecture that
the $\mathcal{Q}$ operator~\eqref{Q_general}
satisfies the $N$-th order difference equation,
\begin{equation}
  \label{Baxter_for_n}
  \mathcal{Q}(\lambda + 2 \, \mathrm{i} \, \gamma)
  -
  \sum_{k=1}^{N-1}
  t_k(\lambda + 2 \, \frac{N-k}{N} \, \mathrm{i} \, \gamma) \cdot
  \mathcal{Q}(\lambda + 2 \,\frac{N-k}{N} \, \mathrm{i} \, \gamma) \,
  +
  \Delta(\lambda+\frac{2}{N}\, \mathrm{i} \, \gamma) \cdot
  \mathcal{Q}(\lambda)
  = 0 ,
\end{equation}
with some function $\Delta(\lambda)$.
Here the transfer matrix $t_1(\lambda)$ is as
eq.~\eqref{NKP_monodromy} and corresponds to the vector
representation.
Other transfer matrices $t_{k>1}(\lambda)$ 
are for $[1^k]$,
and
can be constructed by the fusion method from  the
transfer matrix $t_1(\lambda)$
with a suitable normalization.
Especially we have
\begin{equation*}
  t_{N-1}(\lambda)
  \propto
  \Tr
  \left(
    \prod_n^\curvearrowleft
    \exp
    \Bigl(
    - \sum_{a=1}^{N-1} \mathbf{h}_a \, \chi_n^{(a)} 
    \Bigr) \cdot
    \overline{\mathbf{X}}(\lambda)
  \right)  ,
\end{equation*}
with
$\overline{\mathbf{X}}(\lambda) = 1 - \mathrm{e}^{- \lambda} \,
\mathbf{C}^{-1}$.
The duality of the $\mathcal{Q}$ operator also suggests the dual
difference  equation,
\begin{equation}
  \mathcal{Q}(\lambda + 2 \, \mathrm{i} \, \pi)
  -
  \sum_{k=1}^{N-1}
  u_k(\lambda + 2 \, \frac{N-k}{N} \, \mathrm{i} \, \pi) \cdot
  \mathcal{Q}(\lambda + 2 \,\frac{N-k}{N} \, \mathrm{i} \, \pi) \,
  +
  \Tilde{\Delta}(\lambda+\frac{2}{N} \, \mathrm{i} \, \pi) \cdot
  \mathcal{Q}(\lambda)
  = 0  ,
\end{equation}
where the transfer matrices $u_k(\lambda)$ follows from
$t_k(\lambda)$ under a duality transform~\eqref{duality_transform}.


\section{Concluding Remarks}

We have explicitly 
constructed the Baxter $\mathcal{Q}$ operator for the quantum
discrete Boussinesq equation which has $U_q(\widehat{s\ell}_3)$.
As far as the author knows, the explicit form of the $\mathcal{Q}$
operator has been 
known only for the quantum integrable systems
associated to $s\ell_2$, and our $\mathcal{Q}$ operator is the first
example which solves the third order difference equation.
In our construction,
the $\mathcal{Q}$ operator originates from the
generating function of
the local integrals of  motion for a discrete analogue of the affine
$\widehat{s\ell}_3$ Toda field theory,
and we hope this can be applied for other Lie algebras,
to which some attempts have been made~\cite{RInouKHikam00a}.
Important structure is that
it has a kind of duality,
and that
the $\mathcal{Q}$ operator satisfies the
dual Baxter equation.
It was shown~\cite{FSmir00a}
in the case of the discrete KdV equation that
the dual  Baxter equations
can be interpreted as the duality
between homologies and cohomologies of quantized affine hyper-elliptic
Jacobian from the view point of the algebraic geometry.
It will be interesting to give such interpretation in our case.
In such studies
the separation of variables
(see Ref.~\citen{Sklyan95}  for review, and Ref.~\citen{Skly93} for
$s\ell_3$ case)
will be useful.

Also in $s\ell_2$ case, the universal procedure to derive the Baxter
equation was proposed~\cite{AntonFeigi96a}, where the $\mathcal{Q}$
operator was constructed from the universal $\mathcal{R}$ matrix with
the infinite-dimensional $q$-oscillator representation.
This may  be true for $s\ell_N$ case, but we are not sure how to
relate our $\mathcal{Q}$ operator~\eqref{Q_general}
with the universal $\mathcal{R}$ matrix.

The Baxter equation for $s\ell_2$
 appeared  in studies of the second order  ordinary
differential equations;
the spectral determinants for the Schr{\"o}dinger equation is
identified with the $\mathcal{Q}$ operator for the
CFT~\cite{PDoreRTate99b,JSuzu99a,BazhLukyZamo98c}.
Such correspondence is also studied for the higher order differential
equation~\cite{DoreDunnTate00a,JSuzu00b}.
This ``ODE/IM correspondence'' seems to come from a
similarity between the Stokes multiplier and the Lax matrix, and our
results will help further investigations on these topics.


\section*{Acknowledgement}

The author would like to thank Rei Inoue for stimulating discussions.

\newpage
\appendix
\section{Quantum Dilogarithm Function}
\label{sec:dilog}

We review  a $q$-deformation of the dilogarithm
function~\eqref{define_Phi};
\begin{equation*}
  \Phi_\gamma(\varphi)
  =
  \exp
  \left(
    \int_{\mathbb{R} + \mathrm{i} \, 0}
    \frac{
      \mathrm{e}^{- \mathrm{i} \varphi x}
      }{
      4 \, \sh(\gamma x) \, \sh(\pi x)
      }
    \frac{\mathrm{d}x}{x}
  \right) .
  \tag{\ref{define_Phi}}
\end{equation*}
This integral was introduced by Faddeev~\cite{LFadd95a}, and it
corresponds to a non-compact
analogue of the
$q$-exponential function,
\begin{equation}
  \label{q_exponential}
  S_q(w) = \prod_{n=0}^\infty
  \bigl( 1 + q^{2n+1} \, w \bigr)
  =
  \exp
  \left(
    \sum_{k=1}^\infty
    \frac{(-1)^k \, w^k}{k \, (q^k - q^{-k})} 
  \right) .
\end{equation}

We list properties of the integral~\eqref{define_Phi} in
order~\cite{FaddKashVolk00a}.
\begin{itemize}
\item Duality
  \begin{equation}
    \label{duality_Phi}
    \Phi_{\frac{\pi^2}{\gamma}}(\varphi)
    =
    \Phi_\gamma ( \frac{\gamma}{\pi} \varphi) ,
  \end{equation}

\item Zero points,
  \begin{equation*}
    \text{zeros of $\bigl( \Phi_\gamma(\varphi) \bigr)^{\pm 1}$}
    =
    \Bigl\{
    \mp \mathrm{i} \bigl(
    (2 \, m + 1)  \, \gamma + (2 \, n + 1) \, \pi
    \bigr)
    \Big|
    m , n \in \mathbb{Z}_{\geq 0}
    \Bigr\}
  \end{equation*}

\item Inversion relation,
  \begin{align}
    \Theta_\gamma(\varphi)
    & \equiv
    \Phi_\gamma( \varphi) \cdot \Phi_\gamma( - \varphi)
    \nonumber
    \tag{\ref{introduce_Theta}}
    \\
    & =
    \exp
    \left(
      - \frac{1}{2 \,  \mathrm{i} \, \gamma}
      \left(
        \frac{\varphi^2}{2} +
        \frac{\pi^2 + \gamma^2}{6}
      \right)
    \right) .
    \label{define_Theta}
  \end{align}
  A reason why we use notation ``$\Theta$'' is that
  it is proportional to the
  Jacobi theta function when we replace
  the integral $\Phi_\gamma(\varphi)$ with  $S_q(w)$,
  $S_q(w) \, S_q(w^{-1})$.

\item Difference equations,
  \begin{gather}
    \label{difference_Phi}
    \frac{\Phi_\gamma(\varphi+ \mathrm{i} \, \gamma)}
    {\Phi_\gamma(\varphi - \mathrm{i} \, \gamma)}
    =
    \frac{1}{1 + \mathrm{e}^\varphi} .
  \end{gather}
  The duality~\eqref{duality_Phi} gives another type of difference
  equation,
  \begin{gather}
    \frac{\Phi_\gamma(\varphi+ \mathrm{i} \, \pi)}
    {\Phi_\gamma(\varphi - \mathrm{i} \, \pi)}
    =
    \frac{1}{1 + \mathrm{e}^{\frac{\pi}{\gamma}\varphi}} .
  \end{gather}

\item Pentagon relation~\cite{ChekFock99a,FaddKashVolk00a}
  \begin{equation}
    \label{pentagon}
    \Phi_\gamma(\Hat{p}) \, \Phi_\gamma(\Hat{q})
    =
    \Phi_\gamma(\Hat{q}) \, \Phi_\gamma(\Hat{p} + \Hat{q}) \,
    \Phi_\gamma(\Hat{p}) ,
  \end{equation}
  where $\Hat{p}$ and $\Hat{q}$ are the Heisenberg operators
  satisfying a commutation relation,
  \begin{equation}
    \label{p_q_canonical}
    [ \Hat{p} ~,~ \Hat{q} ] = - 2 \, \mathrm{i} \, \gamma .
  \end{equation}
  This identity is a key to apply the integral~\eqref{define_Phi} to
  the quantization of the Teichm{\"u}ller space~\cite{ChekFock99a} and
  to
  the construction of the invariant of
  3-manifold~\cite{Kasha95,Hikam00a}.
\end{itemize}

\section{Properties of the Fundamental $\mathcal{L}$ Operator}
\label{sec:operator}

We study the property of the fundamental $\mathcal{L}$
operator~\eqref{fundamental_L}.

\subsection{$\mathbb{Z}_3$ Symmetry}

By construction the free fields $\phi_n^{(1)}$, $\phi_n^{(2)}$, and
$\phi_n^{(0)} = - \phi_n^{(1)} - \phi_n^{(2)}$ correspond to roots
$\boldsymbol{\alpha}_1$,
$\boldsymbol{\alpha}_2$,
and
$\boldsymbol{\alpha}_0$.
As the fundamental
$\mathcal{L}$ operator $\mathcal{L}_n(\lambda ; \phi)$
is defined as the intertwining operators for these $\mathbb{Z}_3$
symmetric vertex operators~\eqref{intertwine_vertex},
we can suppose the $\mathcal{L}$ operator itself has the
$\mathbb{Z}_3$ symmetry.
This can be checked by recursive uses of the pentagon
identity~\eqref{pentagon} as
\begin{align*}
  \mathcal{L}_n(\lambda ; \phi)
  & =
  \Bigl(
  \Phi_\gamma(\lambda  - \dphi_n^{(2)}) \,
  \Phi_\gamma(\lambda  - \dphi_n^{(1)}) \,
  \Phi_\gamma(2 \, \lambda  + \dphi_n^{(1)}) \,
  \Phi_\gamma(\lambda  + \dphi_n^{(1)}+ \dphi_n^{(2)})
  \Bigr)^{-1}
  \\
  & \qquad \times
  \Theta_\gamma(\dphi_n^{(2)}) \,
  \Theta_\gamma(\dphi_n^{(1)}) 
  \\
  & =
  \Bigl(
  \Phi_\gamma(\lambda  + \dphi_n^{(1)} + \dphi_n^{(2)}) \,
  \Phi_\gamma(\lambda  - \dphi_n^{(2)}) \,
  \Phi_\gamma(2 \, \lambda  + \dphi_n^{(2)}) \,
  \Phi_\gamma(\lambda  - \dphi_n^{(1)})
  \Bigr)^{-1}
  \\
  & \qquad \times
  \Theta_\gamma(\dphi_n^{(1)} + \dphi_n^{(2)}) \,
  \Theta_\gamma(\dphi_n^{(2)}) 
  \\
  & =
  \Bigl(
  \Phi_\gamma(\lambda  - \dphi_n^{(1)}) \,
  \Phi_\gamma(\lambda  + \dphi_n^{(1)} + \dphi_n^{(2)}) \,
  \Phi_\gamma(2 \, \lambda  - \dphi_n^{(1)} - \dphi_n^{(2)}) \,
  \Phi_\gamma(\lambda  - \dphi_n^{(2)})
  \Bigr)^{-1}
  \\
  & \qquad \times
  \Theta_\gamma(\dphi_n^{(1)}) \,
  \Theta_\gamma(\dphi_n^{(1)} + \dphi_n^{(2)})  .
\end{align*}
As a result, we see that
\begin{align}
  \label{Z_symmetry_F}
  \mathcal{L}_n(\lambda ;\phi)
  \equiv
  \Hat{\mathcal{L}}(\lambda ; \dphi_n^{(1)} ,
  - \dphi_n^{(0)})
  =
  \Hat{\mathcal{L}}(\lambda ; \dphi_n^{(2)} ,
  - \dphi_n^{(1)})
  =
  \Hat{\mathcal{L}}(\lambda ; \dphi_n^{(0)},
  -\dphi_n^{(2)})
  .
\end{align}

\subsection{Unitarity}

Recalling the fact that the $\Theta$ function~\eqref{define_Theta} is
proportional to the Gaussian,
we can check that the fundamental $\mathcal{L}$ operator satisfies the
unitarity condition,
\begin{equation}
  \mathcal{L}_n (\lambda ; \phi) \cdot
  \mathcal{L}_n (-\lambda ; \phi)
  =
  \mathrm{e}^{-\frac{1}{2 \mathrm{i} \gamma} \lambda^2}  .
\end{equation}

\section{Proof of Eqs.~\eqref{L_formula_1} --~\eqref{L_formula_2}}
\label{sec:proof}

We sketch the outline of a proof of eqs.~\eqref{L_formula_1}
--~\eqref{L_formula_2}.

To prove eqs.~\eqref{L_formula_1}, we first compute as follows;
\begin{align}
  \frac{1}{
    \mathcal{L}_n(\lambda + \frac{2}{3} \, \mathrm{i} \, \gamma ; \phi)
    } \cdot
  \mathcal{L}_n^{(1)}(\lambda ; \phi)
  & =
  \frac{\Phi_\gamma(\lambda + \frac{2}{3} \, \mathrm{i} \, \gamma
    - \dphi_n^{(1)} - \dphi_n^{(2)})}
  {\Theta_\gamma(\dphi_n^{(1)} + \dphi_n^{(2)})}
  \cdot
  \frac{\Theta_\gamma(\dphi_n^{(1)} - \frac{2}{3} \, \mathrm{i} \, \gamma)}
  {\Theta_\gamma(\dphi_n^{(1)})}
  \nonumber 
  \\
  & \qquad \times
  \frac{\Phi_\gamma(2 \, \lambda + \dphi_n^{(1)} + \frac{4}{3} \,
    \mathrm{i} \, \gamma)}{
    \Phi_\gamma(2 \, \lambda + \dphi_n^{(1)} - \frac{2}{3} \,
    \mathrm{i} \, \gamma)}
  \cdot
  \frac{\Theta_\gamma(\dphi_n^{(1)}+\dphi_n^{(2)}+ \frac{2}{3} \, \mathrm{i}
    \, \gamma)}{
    \Phi_\gamma(\lambda - \frac{2}{3} \, \mathrm{i} \, \gamma -
    \dphi_n^{(1)} - \dphi_n^{(2)})}
  \nonumber 
  \\
  & =
  \frac{\Phi_\gamma(\lambda + \frac{2}{3} \, \mathrm{i} \, \gamma
    - \dphi_n^{(1)} - \dphi_n^{(2)})}
  {\Theta_\gamma(\dphi_n^{(1)} + \dphi_n^{(2)})}
  \cdot
  \frac{1}{
    1+ \mathrm{e}^{2 \lambda + \frac{1}{3} \mathrm{i} \gamma +
      \dphi_n^{(1)}}
    }
  \nonumber
  \\
  & \qquad \times
  \frac{\Theta_\gamma(\dphi_n^{(1)}+\dphi_n^{(2)}+ \frac{4}{3} \, \mathrm{i}
    \, \gamma)}{
    \Phi_\gamma(\lambda - \frac{4}{3} \, \mathrm{i} \, \gamma -
    \dphi_n^{(1)} - \dphi_n^{(2)})}
  \cdot
  \mathrm{e}^{\frac{1}{3} \dphi_n^{(1)} - \frac{1}{9} \mathrm{i}
    \gamma}
  \nonumber
  \\
  & =
  \frac{1}{
    \mathrm{e}^{\lambda + \frac{2}{3} \mathrm{i} \gamma}
    +
    \mathrm{e}^{2 \lambda + \frac{1}{3} \mathrm{i} \gamma
      + \dphi_n^{(1)}}
    +
    \mathrm{e}^{\dphi_n^{(1)}+\dphi_n^{(2)}+ \mathrm{i} \gamma}
    }
  \nonumber 
  \\
  & \qquad \times
  \frac{
    \Theta_\gamma(\dphi_n^{(1)}+\dphi_n^{(2)} + \frac{4}{3} \, \mathrm{i} \,
    \gamma)
    }{
    \Theta_\gamma(\dphi_n^{(1)}+\dphi_n^{(2)} + 2 \, \mathrm{i} \,
    \gamma)
    }    
  \cdot
  \mathrm{e}^{\frac{1}{3}\dphi_n^{(1)} - \frac{1}{9}\mathrm{i} \gamma}
  \nonumber 
  \\
  & =
  \mathrm{e}^{\frac{1}{3} \dphi_n^{(1)} + \frac{1}{3} \dphi_n^{(2)}}
  \cdot
  \frac{1}{
    \mathrm{e}^{\lambda + \frac{1}{3} \mathrm{i} \gamma}
    +
    \mathrm{e}^{2 \lambda + \frac{2}{3} \mathrm{i} \gamma +
      \dphi_n^{(1)}}
    +
    \mathrm{e}^{\dphi_n^{(1)} + \dphi_n^{(2)}}
    } .
  \label{temporal}
\end{align}
This proves the first equality in eqs.~\eqref{L_formula_1}.
Remaining equalities
follow from
eqs.~\eqref{formula_for_F}.

To prove eqs.~\eqref{L_formula_2}, we shift
parameters in eq.~\eqref{L_formula_1};
\begin{equation}
  \frac{1}{
    \mathcal{L}_n(\lambda - \frac{2}{3} \, \mathrm{i} \, \gamma; \phi)
    }
  \cdot
  \Hat{\mathcal{L}}(\lambda ; \dphi_n^{(1)}+\frac{2}{3} \, \mathrm{i} \,
  \gamma ,
  \dphi_n^{(1)} + \dphi_n^{(2)} - \frac{2}{3} \, \mathrm{i} \,
  \gamma )
  =
  \mathrm{e}^{-\frac{1}{3} \mathrm{i} \gamma}
  \,
  \Bigl(
  c_n^{(1)} \, \mathrm{e}^{2 \lambda}
  + c_n^{(2)}
  + c_n^{(3)} \, \mathrm{e}^{\lambda}
  \Bigr) .
\end{equation}
Using eqs.~\eqref{L_formula_1} and above identity, we can prove the
first equality
in eqs.~\eqref{L_formula_2} as
follows;
\begin{align*}
  \mathcal{L}_n^{(1)} (\lambda ;\phi) \cdot \mathrm{e}^\lambda
  -
  \mathcal{L}_n^{(2)}(\lambda ; \phi)
  & =
  \mathcal{L}_n^{(2)}(\lambda ; \phi)
  \cdot
  \biggl(
  \bigl(
  c_n^{(1)} + c_n^{(2)} \mathrm{e}^\lambda + c_n^{(3)}
  \mathrm{e}^{-\lambda}
  \bigr)
  \cdot
  \frac{\mathrm{e}^\lambda}{
    c_n^{(1)} \mathrm{e}^\lambda + c_n^{(2)} \mathrm{e}^{-\lambda}
    + c_n^{(3)}
    }
  - 1
  \biggr)
  \\
  & =
  \mathcal{L}_n^{(2)}(\lambda ; \phi)
  \cdot
  (\mathrm{e}^{2 \lambda} - \mathrm{e}^{-\lambda})
  \cdot
  c_n^{(2)}
  \cdot
  \frac{1}{
    c_n^{(1)} \mathrm{e}^\lambda
    +
    c_n^{(2)} \mathrm{e}^{-\lambda}
    +
    c_n^{(3)}
    }
  \\
  & =
  (\mathrm{e}^{3 \lambda} - 1 )  \cdot
  c_n^{(2)} \cdot
  \Hat{\mathcal{L}}(\lambda ; \dphi_n^{(1)} + \frac{2}{3} \, \mathrm{i} \,
  \gamma ,
  \dphi_n^{(1)} +   \dphi_n^{(2)} - \frac{2}{3} \, \mathrm{i} \,
  \gamma)
  \\
  & \qquad \times
  \frac{1}{
    c_n^{(1)} \mathrm{e}^{2 \lambda}
    +
    c_n^{(2)} 
    +
    c_n^{(3)} \mathrm{e}^{\lambda}
    }
  \\
  & =
  (\mathrm{e}^{3 \lambda} - 1) \, \mathrm{e}^{-\frac{1}{3} \mathrm{i}
    \gamma}
  \cdot
  c_n^{(2)}
  \cdot
  \mathcal{L}_n(\lambda - \frac{2}{3} \, \mathrm{i} \, \gamma; \phi) 
  .
\end{align*}
Other equalities directly follow  from the $\mathbb{Z}_3$
symmetry~\eqref{Z_symmetry_F}.

\newpage


\end{document}